\documentclass[fleqn,usenatbib]{mnras}

\usepackage{newtxtext,newtxmath}
\usepackage[T1]{fontenc}

\DeclareRobustCommand{\VAN}[3]{#2}
\let\VANthebibliography\thebibliography
\def\thebibliography{\DeclareRobustCommand{\VAN}[3]{##3}\VANthebibliography}

\usepackage{graphicx}	% Including figure files
\usepackage{amsmath}	% Advanced maths commands
\usepackage{hyperref}
\usepackage{array}

\title[Short Polar States]{Short Duration Accretion States of Polars as seen in TESS and ZTF data}

\author[C.~Duffy et al.]{
C.~Duffy,$^{1,2}$\thanks{Contact e-mail: \href{mailto:christopher.duffy@armagh.ac.uk}{christopher.duffy@armagh.ac.uk}}
G.~Ramsay,$^{1}$
Kinwah Wu,$^{3}$
Paul A. Mason,$^{4,5}$
P.~Hakala,$^{6}$
D.~Steeghs,$^{2,7}$
M.A.~Wood$^{8}$
\\ \\
$^{1}$Armagh Observatory and Planetarium, College Hill, Armagh, BT61 9DB, UK\\\
$^{2}$Department of Physics, University of Warwick, Gibbet Hill Road, Coventry, CV4 7AL, UK\\
$^{3}$Mullard Space Science Laboratory, University College London, Holmbury St. Mary, Surrey RH5 6NT, UK\\
$^{4}$New Mexico State University, MSC 3DA, Las Cruces, NM, 88003, USA\\
$^{5}$Picture Rocks Observatory, 1025 S. Solano Dr. Suite D., Las Cruces, NM 88001, USA\\
$^{6}$Finnish Centre for Astronomy with ESO (FINCA), Quantum, FI-20014, University of Turku, Finland\\
$^{7}$OzGrav: The ARC Centre of Excellence for Gravitational Wave Discovery, Clayton VIC 3800, Australia\\
$^{8}$Department of Physics and Astronomy, Texas A\&M University--Commerce, Commerce, TX, 75428, USA\\
}
\date{Accepted 2022 August 31. Received 2022 August 31; in original form 2022 June 27}

\pubyear{2022}

\begin{document}
\label{firstpage}
\pagerange{\pageref{firstpage}--\pageref{lastpage}}
\maketitle

\begin{abstract}
Polars are highly magnetic cataclysmic variables which have been long observed to have both high and low brightness states. The duration of these states has been previously seen to vary from a number of days up to years. Despite this; these states and their physical origin has not been explained in a consistent manner. We present observations of the shortest duration states of a number of Polars observed by ZTF and TESS. This has allowed us to determine that short duration states are a relatively common feature across the population of Polars. Furthermore we have been able to generalise the model of star spot migration to explain both short lived high and low states in Polars by incorporating the interaction between the magnetic field of the white dwarf and that of the star spots.
\end{abstract}


%%%%%%%%%%%%%%%%%%%%%%%%%%%%%%%%%%%%%%%%%%%%%%%%%%

%%%%%%%%%%%%%%%%% BODY OF PAPER %%%%%%%%%%%%%%%%%%

\section{Introduction}

Cataclysmic variables (CVs) are a class of close binary stars consisting of an accreting white dwarf (WD) primary and a late type secondary, or donor, star. In these binaries, matter is transferred from the donor onto the WD via Roche-lobe overflow through the L1 inner Lagrange point. In most cases the accretion flow takes the form of an accretion disc, as matter spirals toward the WD as a result of the viscosity of the disc. These systems often show distinct brightness states, known as high and low states, as a result of differing disc viscosity states and hence mass transfer rates e.g., the Z Cam-like binaries.

The formation of an accretion disc occurs if the WD in the CV is not strongly magnetic. If, on the other hand, the WD has a magnetic field that exceeds $\sim10$ MG, then an accretion disc may not form; instead the matter accretes directly onto regions close to the WD magnetic poles -- the plasma having been channelled by the magnetic field lines since passing through the L1 region. The strong magnetic interaction between the WD and its companion gives rise to a torque which forces the binary orbit, the spin of the WD and the other components in the system into synchronous rotation. Such systems are known as Polars or AM Her systems after the archetypal source AM Herculis. See \citet{warner_1995} for a review of CVs in general and \citet{1990SSRv...54..195C,2000SSRv...93..611W} for Polars and their magnetic-field channelled accretion in particular.

Whilst about 140 Polars are now known, the mechanism(s) responsible for shutting off and on the mass transfer through L1 remain poorly understood. The Catalina Real-Time Transient Survey observed 44 Polars over a duration of 9 years \citep{2015gacv.workE..16M}. In that study, Polars identified to have short-lived low states included V1309 Ori, V834 Cen, and EQ Cet. Those undergoing short-lived high states included AR UMa, EF Eri, and HS Cam, whilst a few Polars -- FL Cet, ST LMi, and BM CrB -- have three brightness states. As they lack of an accretion disc, Polars are therefore excellent objects to study how the process of mass transfer can switch off and on in accreting binaries. 

%Regardless of its brightness state, accretion flows in Polars moves along the magnetic field lines to impact the WD at the footpoint of a particular field line. 
A second reason to study polars is that they are ideal objects to study how accreting material in general interacts with strong magnetic fields. The magnetic field lines which the matter threads onto are determined by the ram pressure of the accretion flow. During a high rate of accretion the ram pressure is high and the material can penetrate deeper into the WD magnetic field than under a lower accretion rate. This allows the material to ``latch onto'' different field lines and potentially accrete onto different or additional footpoints on the WD surface. Such changes in accretion geometry can be identified by changes to the orbital lightcurve of a system. In addition to the presence of different multiple or different accretion regions the observed emission from these regions can vary; it was proposed by \citet{1982A&A...114L...4K} that accretion can proceed via a method termed ``blobby accretion''. Under this model individual ``blobs'' of material accrete along the magnetic field lines as opposed to a continuous stream. This model has been successful in explaining differences in accretion at different poles \citep{1988MNRAS.235..433H,2020A&A...642A.134S}.

In addition to Polars state transitions have been observed in Intermediate Polars (IPs) and VY Scl stars. Unlike Polars these types of CVs have accretion discs to some degree and whilst this brings additional considerations to the mass transfer properties, it offers us an additional population of objects to aid us gain insight into state transitions in Polars. IPs are magnetic CVs with a magnetic field $<10$MG, as such unlike their more strongly magnetic cousins they do form a (truncated) accretion disc. Recently \citet{2022ApJ...928..164C} investigated the low states in IPs. Of the systems which they investigated all but one exhibited long-lived (of order months and years) low states. This implies that IPs are predisposed to more long lasting and stable state changes than those often seen in Polars. Some IPs are thought to be, at least at times, disc-less, and at least in the case of YY Dra (DO Dra) can have short-lived low states \citep{2022AJ....163..246H}. This suggests that mechanisms behind low state formation in stream accreting IPs are different than for the usual IPs with discs. For example, the disc may have a screening effect for the WD magnetic field at L1 as well as acting as a buffer between changes in mass transfer through L1 and the observed accretion rate onto the WD.

The archetype Polar, AM Her, displays frequent state transitions and as such has been the subject of most of the investigations into state transitions in Polars, and hence is a useful object to consider at length. AM Her was the first Polar to be identified, because of its brightness (high state brightness $\sim$13.5 mag); and its strongly polarised optical radiation was discovered to vary with the binary orbital period of 185.6 min \citep{Tapia77}. AM Her has been studied at length, with optical data going back over 100 years. As such the long term macro behaviour of AM Her has been extensively studied \citep[e.g.][]{1995MNRAS.276.1382R,2008A&A...481..433W,2016MNRAS.463.1342S}. In particular, AM Her is known to exhibit two different brightness states; a high and low state. As opposed to many other CVs which have different brightness states which they transition between with a singularly defined recurrence time and duration of states, AM Her shows a mixture of long and short, low and high, states which can last from several days to months for low states, and several months to years in the case of high states, with as-yet no discernible overall pattern. Consequently, it has been a challenge to quantify the long term behaviour of AM Her. Efforts to this end have identified that there are a number of different time-scales upon which these different states occur and have verified that these time-scales and state duty cycles can be determined, however their presence is intermittent \citep{2008A&A...481..433W,2016MNRAS.463.1342S}.

\citet{1991MNRAS.251...28W} found that during the high state of AM Her, the material accreted onto two distinct regions on the WD; a primary and a secondary. The primary region is believed to occupy part of the face opposite the donor and accounts for the majority of the mass transfer and emission \citep[see Figure~18 of][]{1991MNRAS.251...28W}. The secondary accretion region is found on the donor star facing surface of the WD, however at a separation substantially less that $180\degr$ from the primary. This fact excludes the simplest models for explaining the behaviour seen in AM Her, namely that of centred dipole \citep{1993MNRAS.260..141W}. This has been reinforced by observations of other Polars which indicate that the primary accretion region has a magnetic field $\sim2$ orders of magnitude weaker than that found at the secondary region \citep{1989ApJ...337..832F,1989ApJ...342L..35W,1990A&A...230..120S}. \citet{1991MNRAS.253P..11W} presented a multipole model comprising of a dipole and quardupole in the WD and an intrinsic dipole in the donor star and further elaborated it in \citet{1993MNRAS.260..141W} to explain the observations of accretion geometry and distribution of magnetic field strengths between the accretion regions. Whilst the model provided an explanation to the general accretion behaviour of Polars, and AM Her in particular, the precise location of the accretion region remains a subject for more thorough investigations. Efforts to map the cyclotron emission region have some gained success in a number of systems, e.g., CD Ind \citep{2019MNRAS.486.2549H}.

%\textit{EXOSAT} observations of AM Her \citep{1985A&A...148L..14H} indicated that the brightest soft X-ray emissions had unexpectedly moved to originate from the secondary accretion region whilst the hard X-rays had remained present at the primary region. \citeauthor{1985A&A...148L..14H} attributed it to the change in the mass transfer rates in the accretion onto the two poles, creating a ``reversed mode''. A followup study by \citet{1986MNRAS.221..513M} considered the optical lightcurves of AM Her during this reversed mode, and found no change in the optical lightcurve except a slight decrease in overall system brightness. X-ray observations from 2015 of the reversed accretion mode found that the onset of this mode was not triggered by significant changes in the accretion rate \citep{2020A&A...642A.134S}.

In the low state, with reduced mass transfer, AM Her was observed to accrete onto only one pole. \citet{2008A&A...481..433W} suggested that this transition into a low state was due to a realignment of the system magnetic field with the WD being the key component regulating mass transfer. This is an extension of an earlier proposition by \citet{1994ApJ...427..956L}, in which the suppression of the mass transfer that leads to the low state is caused mainly by the migration of a star spot or several star spots on the donor star to regions close to the L1 point of the binary. While promising, this scenario needs further development so as to provide a self-consistent explanation for the duration and frequency of the low states observed in some systems. More specifically, it must explain how the secondary stars, which are low-mass M stars, generate a sufficient quantity of star spots and what mechanisms entice the star spots to migrate to regions sufficiently close to the L1 point such that the mass transfer would be efficiently disrupted \citep[see e.g.][]{2000A&A...361..952H}. Nonetheless, the study of \citet{2005AJ....130..742K} identified that the timescale of the transition into a low state in AM Her, and other Polar above the period gap, was consistent with that expected of the star spot scenario.

In this work we seek to further understand the mechanism which controls these state changes and explain both the short duration high and low states in a consistent fashion. To that end we present observations of a number of short duration states in Polars in both the {{Zwicky Transient Facility (ZTF)}} and the {{Transiting Exoplanet Survey Satellite (TESS)}}. We use these observations to probe the properties of these state changes and to develop an understanding of the changes in accretion geometry which are associated with them.

\section{Photometric Data}

In looking for short duration state changes among Polars we focused on two different data sets: ZTF and TESS. ZTF is a time domain survey program making using of the Palomar 1.2m telescope; covering $47 \deg^2$ per exposure ZTF is able to survey large areas of the sky rapidly building up lightcurves of diversely located sources to a limiting brightness of $\sim 20$ mag. Observations are made in three different filters, ZTF-\textit{i}, ZTF-\textit{g}, ZTF-\textit{r} although in this work we limit our analysis to the \textit{g} and \textit{r} bands ($\sim4087-5522$\AA\ and $\sim5600-7316$\AA\ respectively) as the observing program employed by ZTF results in these filters having the highest cadence \citep[see][for a full description]{2019PASP..131a8002B}.

TESS is a space based optical observatory launched in 2018. It has an observing strategy of making CCD observations with 4 wide-field telescopes, each with a $24\degr\times24\degr$ field of view. The observing plan splits the sky into 26 set sectors spanning from the ecliptic to the ecliptic poles. Each of these sectors are observed continuously for approximately 27 days with a short gap at the midpoint for data transmission and momentum unloading. During an observation, 120 second cadence photometry is generated on a series of predefined targets \citep[see][for full TESS details]{2015JATIS...1a4003R}\footnote{The sources in this paper were included on the 2 min cadence list thanks to their inclusion on the following Guest Investigator programs: GO11256/PI Garnavich; G011268/PI Scaringi; G022071/PI Scaringi; GO22116/PI Wood; GO22230/PI Littlefield; GO3245/PI Littlefield; GO3071/PI Scaringi; GO4009/PI Ramsay; GO4046/PI Scaringi; GO4208/PI Littlefield}.

\begin{figure*}
\begin{center}
\includegraphics[width=\textwidth]{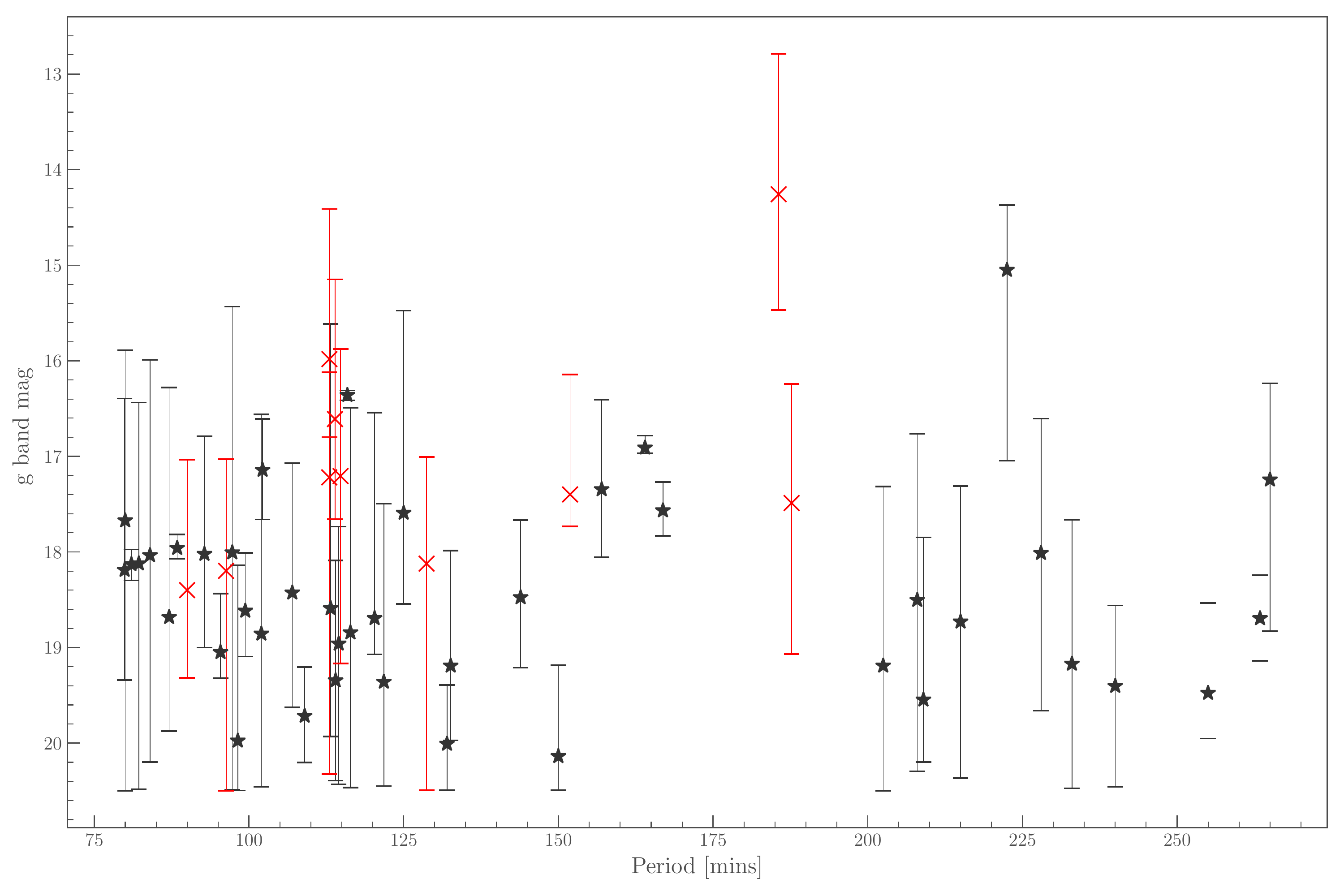}
\caption{Period distribution of the Polars contained within the ZTF sample considered as the mean \textit{g} band magnitude as a function of period. Error bars show the max-min range of the brightness. Entries which are shown in grey with a $\star$ denote those systems which do not show short transitions; entries in shown in red with an X denote those which do show transitions. Full listing of these sources is contained in \autoref{appx1}.}
\label{ZTFPopDistro}
\end{center}
\end{figure*}

\begin{figure*}
\begin{center}
\includegraphics[width=\textwidth]{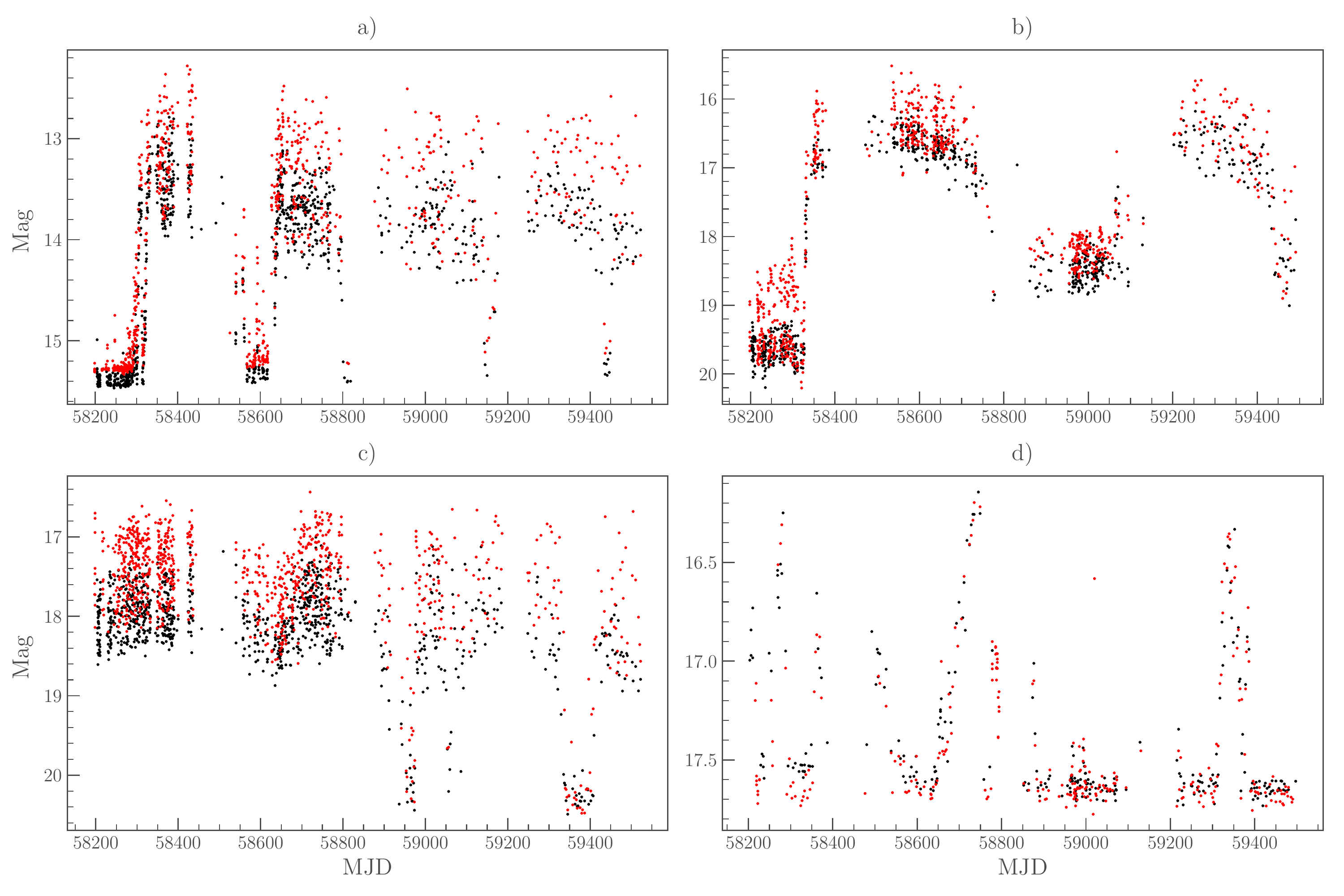}
\caption{Lightcurves of 4 of the objects considered with ZTF data coloured red and black to denote ZTF-\textit{r} and ZTF-\textit{g} respectively showing a) AM Her with both long and short duration states, b) SDSSJ154104+360252 which shows only long duration state changes, c) MT Dra which shows only short duration state changes, and d) AP CrB which shows only short state changes to a higher state}\label{ZTfPlots}
\end{center}
\end{figure*}

\subsection{ZTF Data}\label{sec:ztfData}
We accessed photometric data of 60 known Polars \citep[private communication]{Ramsay} collected by ZTF (see \autoref{appx1} for full details). Of these, 55 sources were deemed to have enough data for analysis; i.e., at least 25 data points in both ZTF-\textit{g} and ZTF-\textit{r}. The distribution of this sample is illustrated in \autoref{ZTFPopDistro} which shows the mean \textit{g} band magnitude of the sources as a function of their period, with error bars used to denote the magnitude range of the system. This distribution clearly shows the period gap, below which the majority of Polars are found. We see that systems below the period gap show a greater spread in magnitude range with systems that vary by as little as $\sim 0.2$ mag to as much as $\sim 4$ mag whereas those systems above the period gap have a more consistent magnitude spread of $\sim 2.4$ mag.

These sources were first visually inspected for brightness state changes. For those systems exhibiting a state change, we performed subsequent analysis. This inspection revealed that within the state changing systems there were three categories of behaviour; systems which showed ``long'' lived state changes, systems which showed ``short'' lived state changes and systems which showed both. Examples of systems which show each such behaviour are shown in \autoref{ZTfPlots}, which also shows an example of ``short'' lived state changes where brightness increases as well as those where it decreases, the full listing of behaviours can be found in \autoref{appx1}.

We established a definition of ``short'' lived state changes to be those which had a duration, the time between onset and the return on the original brightness, of $\lesssim 60$ days. This was established via a combination of inspection of the population of lightcurves and comparison with previous studies by \citet{2005AJ....130..742K}.

We employed two methods to establish the duration of states depending on density and temporal distribution of the data available. Whenever the data sampled the short duration state and its transition sufficiently well, we undertook a non-linear curve fitting approach. Making use of the Python module \texttt{lmfit} \citep{matt_newville_2021_5570790} we defined the short lived state as a composite of a skewed Voigt profile; to fit the short duration state, and a constant function; to fit to the proceeding and succeeding state. As a convolution of the Gaussian and Lorentz profiles the Voigt profile comprises of two parameters which control the width of the curve; $\sigma$ and $\gamma$. In initial tests using this model we set these parameters to be independent but we found that they both converged on the same value, as such the analysis was carried out under the assumption that they were of equal value.

We performed our minimisations using a Markov chain Monte Carlo (MCMC) algorithm making use of the \texttt{emcee} \citep{2013PASP..125..306F} Python module to do so. Although this increased the analysis compute time compared to a least squares minimisation, it improved the robustness of our fits by allowing us to visualise the parameter space and to minimise and isolate uncertainty in individual fitted parameters in order to improve our confidence in a given fit. 

\begin{figure}
    \begin{center}
    \includegraphics[width=\columnwidth]{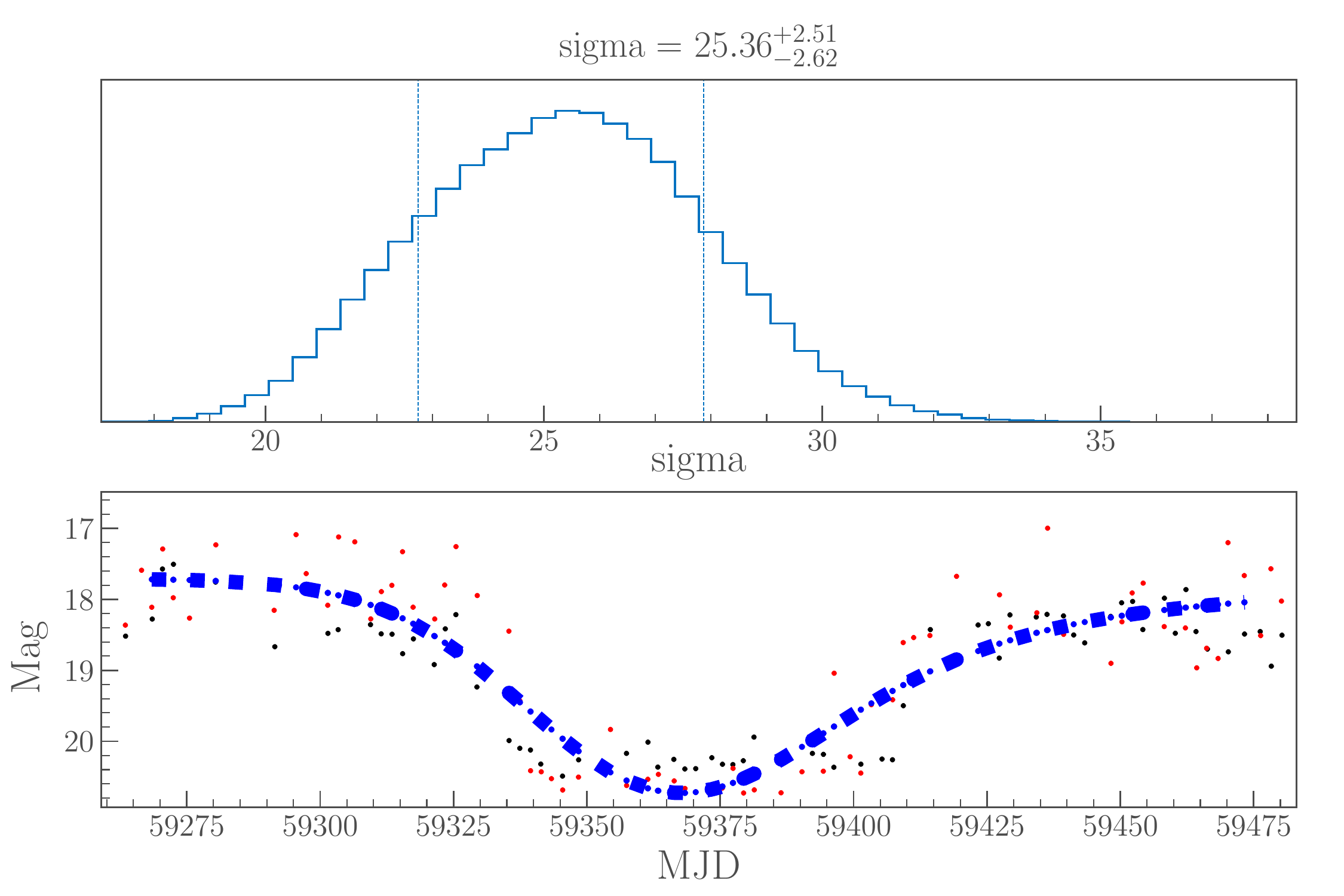}
    \caption{Upper: the result of the marginalisation of the sigma parameter, extracted from the multiparameter analysis for a state transition in MT Dra. Lower: the fitted state transition in MT Dra and the fitted data. Red data points indicate \textit{r} band and black data points indicate \textit{g} band data.}
    \label{fig:exampleFit}
    \end{center}
\end{figure}

We established the duration of the fitted short-lived states from the full width half maximum (or minimum) value for the Voigt profile. In a Voigt profile this is approximated in a relationship with the FWHM values, $F$, of the Gaussian component, \autoref{gauss_FWHM}, and the Lorentzian, \autoref{lorenz_FWHM}: 
%\begin{equation} 
%\label{gauss_FWHM}
%\text{F}_{\mathrm {G} }=2\sigma {\sqrt {2\ln(2)}}
%\end{equation}
%\begin{equation}
%\label{lorenz_FWHM}
%\text{F}_{\mathrm {L} }=2\gamma, 
%\qquad\qquad \begin{aligned} \gamma \equiv \sigma %\end{aligned}
%\end{equation}  
\begin{align} 
\label{gauss_FWHM}
\text{F}_{\mathrm {G} } 
  & =2\sigma {\sqrt {2\ln(2)}}\; \\ 
\label{lorenz_FWHM}
\text{F}_{\mathrm {L} }& =2\gamma  \ , 
\end{align}  
with   $\gamma \equiv \sigma$.  
These are combined to give the following approximation of the FWHM of the Voigt profile which is accurate to $\sim\pm0.02\%$ \citep{1977JQSRT..17..233O}:
\begin{equation}
\text{F}_\mathrm{V}\approx 3.60\sigma \ .
\end{equation}

\autoref{fig:exampleFit} shows the 1D marginalised posteriors of the value of sigma extracted from the results of the MCMC calculation on the fit of a state transition seen in MT Dra, which is shown in the lower panel. This indicates the quality of the fit, particularly with respect to the parameter key to the duration determination; furthermore it illustrates the applicability of the model used to fit these transitions.

The calculated duration of all of these states is contained within \autoref{tab:ZTFsub}. When more than one short-lived state was observed, the mean duration value is shown. In those cases where the data did not permit us to undertake a curve fitting approach, we undertook a visual inspection. This does not allow for the same level of precision available in cases where model fitting is possible, but it ensures that sources which do show these features are not neglected. By combining these methods, the systems which were found to meet these criteria are listed in \autoref{tab:ZTFsub} and highlighted in \autoref{ZTFPopDistro} where they are marked with an X and coloured red.

In \autoref{tab:ZTFsub}, four sources have ancillary values quoted in brackets for some of their parameters. In the case of {ST LMi}, 5 short-lived states are observed; however we were only able to resolve 3 in order to calculate the duration. Despite this, all 5 were used to calculate the mean recurrence time. In each {MT Dra} and {AP CrB} we observed an additional short state which took the same profile as the others seen in the respective lightcurves but had a duration longer than our definition for a short duration event (69.1 and 93.4 days). For these sources, the values contained within brackets relate to the inclusion of these events. Finally, in {AM Her} we observed two events within quick succession, $\sim 2$ days, individually these events would fit in our criteria, however if they are in actual fact a single event then it would not and we cannot with absolute certainty say that these are separate events. As such we have included these two events and their related parameters as ancillary values.

\begin{table}
\caption{ZTF sources which show short-lived state {{changes}}. Showing the number of events their mean duration and mean recurrence. Mean durations marked with $\dagger$ are derived partially from visual inspection, $\dagger\dagger$ denotes mean durations derived entirely from visual inspection. Values contained within brackets are ancillary values calculated on the basis explained {{in \textsection\ref{sec:ztfData}}}. Errors arise from fit confidence values or the standard deviation where visual inspection was used. Full details of sources can be found in \autoref{appx1}.}
\label{tab:ZTFsub}
\begin{tabular}{lcccc}
\hline
Source&\textit{g}\textsubscript{mean}&States&Duration [d]&Recurrence [d]\\
\hline
FL Cet & 19.15 & 4 & $40.9^{\dagger\dagger}\pm10.7$& $180\pm158$\\
AN UMa & 17.21 & 1 & $53.1^{+5.8}_{-4.8}$ &———\\
RXJ1610+03 & 17.49 & 1 & $40.1^{\dagger\dagger}$&———\\
EU UMa & 18.40 & 1 & $39.9^{+3.7}_{-3.9}$ &———\\
ST LMi & 16.61 & 3 (5) & $24.1^{\dagger\dagger}\pm7.6$& $273\pm90$\\
MT Dra & 18.12 & 2 (3) &$27.5^{\dagger}\pm8.9$ (49.5)&110 ($196$)\\
AM Her & 14.26 & 2 (4) & $21.7^{\dagger}\pm7.3$ (27.6) &$297$ ($288$)\\
AP CrB & 17.40 & 4 (5) & $33.9^{\dagger}\pm15.6$ (42.7) &$368\pm201$(276)\\
V2301 Oph & 17.22 & 1 & $37.4^{+5.7}_{-4.1}$&———\\
BS Tri & 18.20 & 1 & $19.1^{\dagger\dagger}$&———\\
V884 Her & 15.98 & 3 &$49.6^{\dagger}\pm6.1$& $389\pm40$\\
\hline
\end{tabular}
\end{table}

We identified 11 sources which show short-lived state changes. This accounts for 18.3\% of the original sample considered. Of the systems identified it can be seen clearly in \autoref{ZTFPopDistro} that they disproportionately, albeit not exclusively, lie below the period gap. Within this distribution there is a particular bunching of 4 sources with periods of $\sim113$ minutes. Overall the mean duration of short-lived state transitions observed was 35.2 days, increasing to 38.5 days if ancillary values in the three cases discussed above are included. Similarly, the mean recurrence time of those systems which show multiple events is 270 days falling only slightly to 267 days when the ancillary values are included. No discernible relationship appears to exist connecting recurrence time or duration to either the orbital period or the $B$ field strength. We furthermore performed an Anderson-Darling test on the period distributions the sources with and without short duration states; this yielded no evidence of a significant difference in the distribution of each population.

\subsection{TESS}
We accessed TESS two minute photometry on 9 Polars through the Python module \texttt{lightkurve} \citep{2018ascl.soft12013L}. The high temporal resolution offered by TESS allowed us to both study the short-lived states in more detail, which had been identified in the ZTF data and probe sources for short duration events that could not otherwise be resolved. \autoref{tab:TESSSamp} summarises the sources studied and identifies those in which a short state occurred.

\begin{figure*}
\begin{center}
\includegraphics[width=\textwidth]{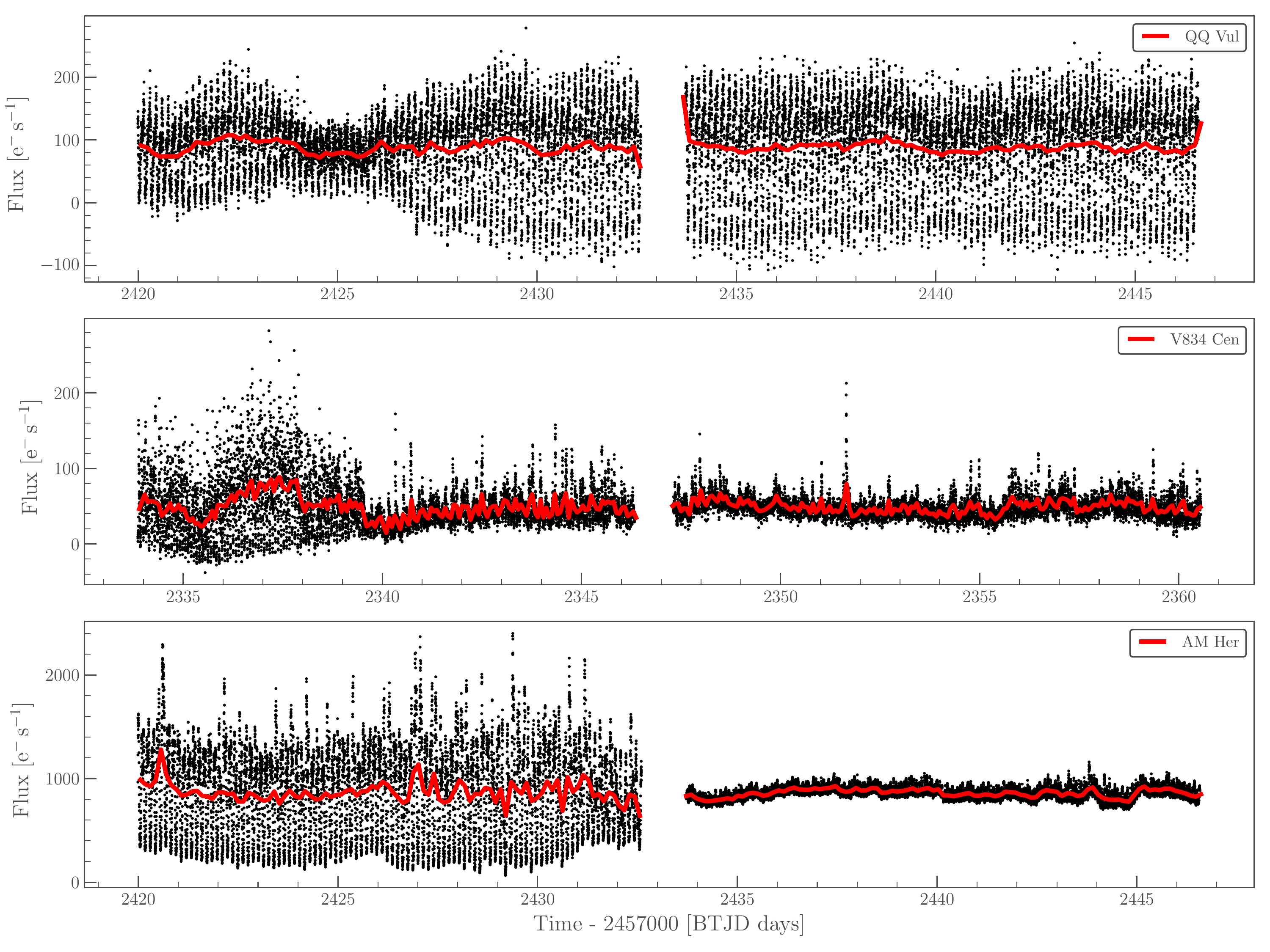}
\caption{TESS lightcurve of a) QQ Vul sector 41, b) V834 Cen sector 38, and c) AM Her sector 41. In each panel, the lightcurve binned on the known orbital period of the source is shown in red.}
\label{fig:SampleTESS}
\end{center}
\end{figure*}

The events which we identified fall into three broad categories; short-lived dimming and brightening events; both of which have a duration $\lesssim5$ days. The third category are the counterparts of the events which had previously been identified from the ZTF data. \autoref{fig:SampleTESS} show examples of each of these, where the TESS lightcurve is shown in black with the same binned upon the known orbital period superimposed in red. In total we identified 5 sources with one of these behaviours. Using these findings from TESS we are able to observe how the orbital lightcurve of these sources change as a result of moving into and out of these states (see \textsection\ref{sec:profile}). We can then use this to infer how, if at all, the accretion geometry has changed as a result of the state change, which could help us determine how the accretion flow attaches onto different magnetic field lines depending on the mass transfer rate or other factors. 

\begin{table*}
\caption{Properties of the TESS sources considered and the sectors in which they were observed by TESS. Quoted magnitudes are sourced from Gaia eDR3.}\label{tab:TESSSamp}
\begin{tabular}{lcccccl}
\hline
Source & Sectors & Mean G mag & Period [min] & Inclination&Transition & Notes\\
\hline
BL Hyi & 1, 2, 29 & $19.1$ &$123.6^{a}$&$32\degr^{b}$&&  \\
CW Hyi & 1, 2, 27-29& $17.5$ &$181.8^{c}$&$45\degr-85\degr^{d}$& \checkmark & 2 $\sim3$d reductions in brightness\\
FL Cet & 4, 42, 43 & 18.6 &$87^{e}$&--  \\
MR Ser & 24, 25 & $16.1$ &$112.3^{f}$& $45\degr^{g}$  \\
MT Dra & 14-16, 18-26, 40, 41 & 17.5 &$128.7^{h}$&$\leq70\degr^{h}$& \checkmark &Contemporaneous observations of ZTF state\\
UW Pic & 5, 6, 31-33 & $15.9$ &$132.5^{i}$ &$45\degr-75\degr^{i}$&&\\
AM Her & 14, 25, 26, 40, 41 & 14.3 &$185.6^{j}$&$35\degr-60\degr^{k}$& \checkmark & Contemporaneous observations of ZTF state\\
V834 Cen & 38 & $16.6$ &$101.5^{l}$&$45\degr-60\degr^{m}$& \checkmark &$\sim4$d increase in brightness\\
QQ Vul & 41 & 15.3 &$222.5^{n}$&$\sim50\degr^{o}$& \checkmark & $\sim2$d reduction in brightness\\
\hline
\end{tabular}\\
\begin{flushleft}
\textit{Citations}: \footnotesize{$^{a}$\citet{1983Natur.301..318A},$^{b}$\citet{2007A&A...463..647B}, $^{c}$\citet{2002A&A...396..895S}, $^{d}$\citet{1997A&A...327..183B}, $^{e}$\citet{2002AJ....123..430S}, $^{f}$\citet{1991A&A...244..373S}, $^{g}$\citet{BrainerdLamb}, $^{h}$\citet{2002A&A...392..505S}, $^{i}$\citet{2003MNRAS.339..685R}, $^{j}$\citet{2005AJ....130.2852K}, $^{k}$\citet{2020A&A...642A.134S}, $^{l}$\citet{1983ApJ...264..575M}, $^{m}$\citet{1993A&A...267..103S},  $^{n}$\citet{1984ApJ...277..682N}, $^{o}$\citet{1998MNRAS.295..353C}}
\end{flushleft}
\end{table*}

Although the number of systems identified in TESS data to show short duration outbursts is small and cannot be used to draw population-wide conclusions, we note that those systems with the shortest events visible only in TESS appear to be biased toward longer period systems. However as we identified no relationship connecting state duration to the orbital period, we consider that this is most likely simply the result of small number statistics.

\subsection{Orbital Profile Lightcurves}\label{sec:profile}

In each of the sources in \autoref{tab:TESSSamp} for which we found to contain short-lived state changes, we used the TESS data to investigate how the lightcurve phased upon the orbital period (henceforth the orbital lightcurve) evolved as the sources moved between their states. In each case we extracted a portion of the TESS lightcurve which corresponded to the short duration state which we subsequently folded upon the known orbital period and then binned. This process was then repeated with other segments of similar duration from the lightcurve outwith the short-lived state. This allowed us to compare the orbital lightcurves in various states and to identify any differences; in this manner it is possible to constrain the accretion geometry of each state and to understand how it evolves.

Typically, in Polars accretion is generally described as being either one or two pole. In one-pole accretion the accretion region is behind the WD not always visible to us; whereas in two pole systems the orientation of the system is such that one of the accretion region is always visible as the system rotates. Each of these scenarios creates a different orbital lightcurve signature \citep{1990SSRv...54..195C}. As the mass transfer rate in a system changes, the ram pressure of the flow changes accordingly, so material can become captured by different sets of field lines which can result in matter accreting onto different regions of the WD. As such, this can result in a change in the observed orbital lightcurve and provides insight towards the underlying accretion geometry. 

%{\textbf{Some changes will probably be needing to made to each of these sections in general after considering the change in accretion in \textsection3.3, so that that discussion doesn't come entirely out of the blue }}

%\begin{figure}
%\begin{center}
%\includegraphics[width=\columnwidth]{Figures/Spot_dip_outburst.pdf}
%\caption{A decrease in brightness in TESS data of GG Leo is followed by a huge outburst. It is likely that a star-spot moving across L1 %is responsible for the brightness dip, which is interrupted by a magnetically induced short-lived high state.}
%\label{fig:}
%\end{center}
%\end{figure}

\subsubsection{QQ Vul}

\begin{figure}
\begin{center}
\includegraphics[width=\columnwidth]{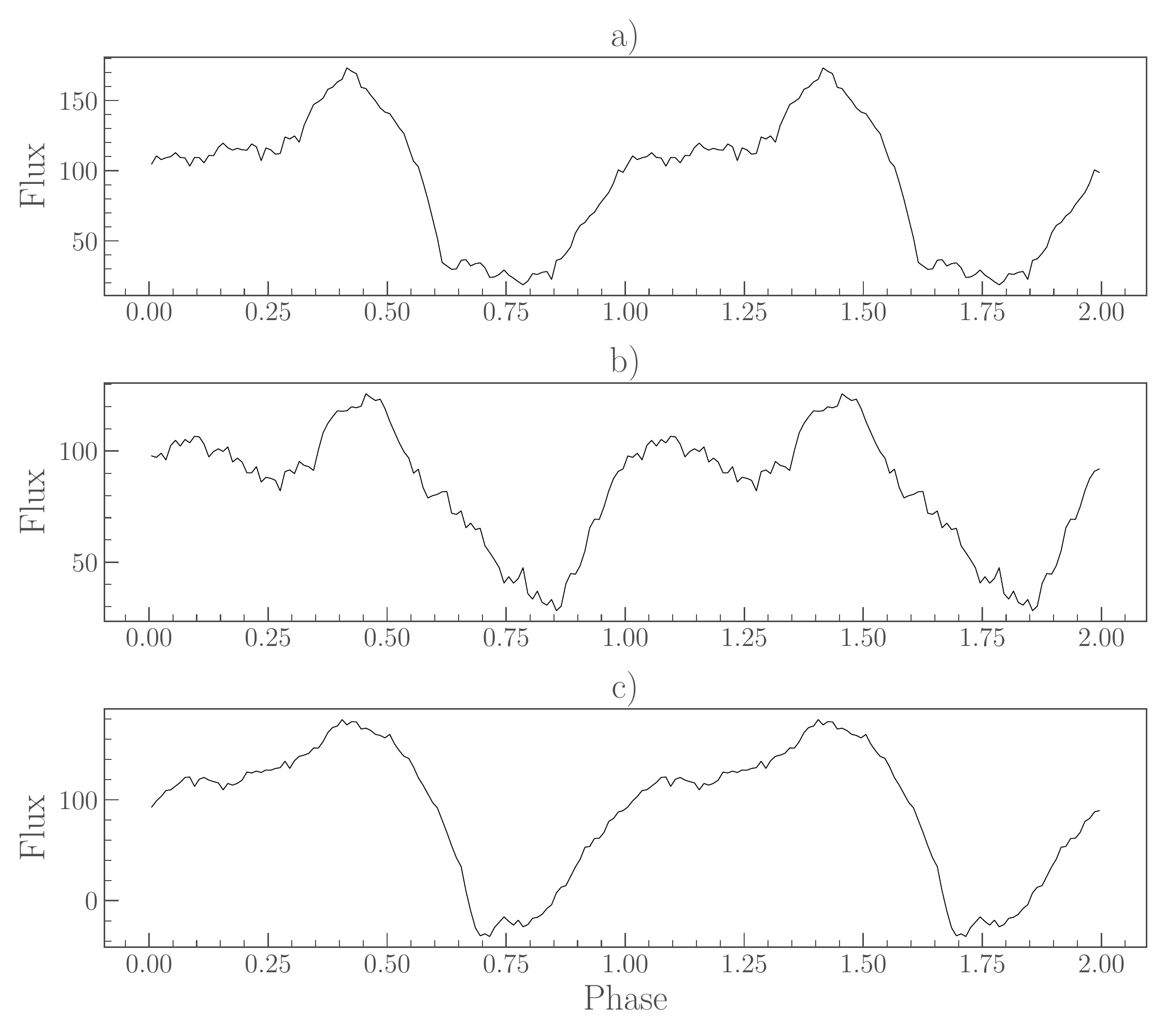}
\caption{Folded and binned lightcurve of {QQ Vul} a) before the decreased brightness event ({BTJD=}2420-2423.5), b) during decreased brightness event ({BTJD=}2423.5-2426.5), and c) after the decreased brightness event ({BTJD=}2426.5-2430.5). All three have been phased with an epoch set to the start of sector 41.}
\label{fig:foldingQQvul}
\end{center}
\end{figure}

 %Unsurprisingly the lightcurve before and after the dimming Event is broadly similar; initially following a mostly flat profile from phase=0.00-0.25 after which brightness increases to a maximum at phase=0.44 before dropping to a minima at phase=0.80 and phase=0.70 before and after the event respectively. In both cases the brightness then begins to increase again. During the decrease in brightness the lightcurve is, for the most part, very similar; excepting the expected change in brightness and the scale of the changes. The most striking difference in the lightcurve during the event is that as opposed to the steady state brightness between phase=0.00-0.25 seen before and after brightness instead decreases by $\sim15\%$ before rising to a maximum value at phase=0.48.

%{\textbf{I think that we need to discuss why the orbital minima substantially increases in brightness during the QQ Vul event, while the mean brightness drops a bit. It seems possible to me that a new, non-self-eclipsing, pole is active during the event. -PM}}

In \autoref{fig:foldingQQvul} we show the folded and binned orbital lightcurve of {QQ Vul} before, during, and after the short duration reduction in brightness. The orbital lightcurves before and after the dip are unsurprisingly similar, indicating that the initial accretion state is reestablished after the dip. The high state orbital lightcurve is similar to the ``Type 1'' state identified by \citet{2003AJ....125.2188K}, where there's only an optical signature from 1 pole. Conversely the lightcurve during the dip is similar to the ``Type 2'' state, which shows an optical signature from a second pole. This dip event lasts $\sim$2 days which is slightly shorter than the finding of \citet{2003AJ....125.2188K} who note they usually persists for more than 3 days. 

\citet{1998MNRAS.295..353C}, \citet{2000MNRAS.313..533S}, and \citet{2003AJ....125.2188K} indicated that {QQ Vul} may be a two pole accretor. X-ray observations made by \citet{2000PASP..112..343B}, which coincided with a ``Type 1'' state, optical observations were shown to originate from two pole accretion. This indicates that this configuration has two pole accretion but that the second pole does not leave an optical signature. Whilst the mean brightness during the dip understandably decreases; the orbital minimum at this time is actually brighter. This suggests that a new accretion footpoint is established at this time.

%It is possible that the optical radiation of the accretion stream onto the primary pole in the high state is such to occlude any signature of accretion onto a second pole. Thus, when {QQ Vul} moves into a lower accretion state, as we see here, the signature of the second pole is revealed as the ``Type 2'' lightcurve. This suggested change in mass transfer rate appears not to change the field lines upon which matter flows along.

\subsubsection{V834 Cen}

\begin{figure}
\begin{center}
\includegraphics[width=\columnwidth]{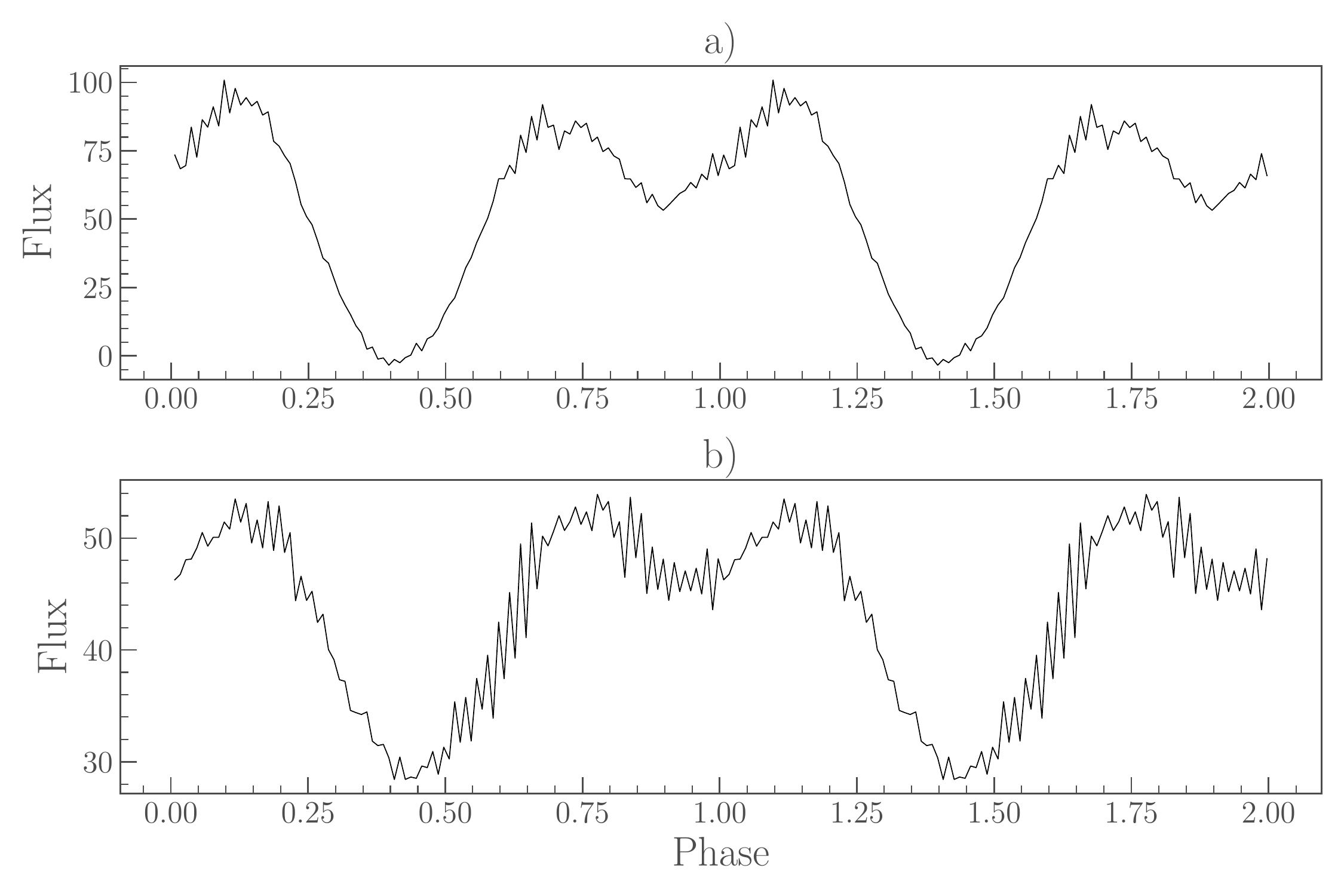}
\caption{Folded and binned lightcurve of {V834 Cen} a) during the increased brightness event ({BTJD=}2334-2339.6) and b) after the increased brightness event ({BTJD=}2339.6-2346.3). Both have been phased with an epoch set to the start of sector 38.}
\label{fig:foldingV834}
\end{center}
\end{figure}

In \autoref{fig:foldingV834} we present the folded and binned orbital lightcurve of {V834 Cen} during the short enhanced brightness event which occurs at the start of the sector 38 observations and the profile immediately after this event. Unlike {QQ Vul}, the two lightcurves shown here are essentially identical, notwithstanding the scale of the brightness changes. This implies that the accretion geometry remains unchanged, and that matter accretes along the same field lines in both. The orbital lightcurve presented here is consistent with the earlier findings of \citet[cf. figure 8]{1986MNRAS.218..201C}. That work also presents a different orbital lightcurve (cf. figure 9) which is suggested as part of the evidence that the accretion stream's location is not fixed. As we do not see this difference between the brightening event and the subsequent dimmer state, we are confident that the accretion stream has a fixed location between these two states.

%Both lightcurves rise slightly until phase=0.25 before falling to a minimum with the lightcurve of the Event achieving minima slightly earlier in the orbital cycle phase=0.47 vs. phase=0.52. Both lightcurves then increase in brightness again with both reaching the local maxima at phase=0.85 before subsequently dimming again. The maxima in the Event are not of equal brightness as the second maxima is $\sim8.7\%$ dimmer than the first; this contrasts to the lightcurve after the event where both maxima are equivalent. 

In the state which follows the event, we see between phase=0.50-0.70 an increase in the {{root mean squared deviation}} of the orbital lightcurve. Comparing with the unfolded lightcurve in the middle panel of \autoref{fig:SampleTESS} we see that this is associated with repeated flare-like events that occur during this state. These events reoccur at approximately the same phase causing them to appear as they do on orbital lightcurve. Similar events have been seen by \citet{1991ApJ...382..315M} and \citet{2017A&A...600A..53M} occurring at the same orbital phase. The origin of these  events has not yet been determined. However, these ``flares'' disappear during the enhanced accretion event, which may be associated with the colour dependency of the flares at different mass transfer rates as identified by \citet{1991ApJ...382..315M}, however given the broad bandpass in TESS this seems unlikely. A more plausible explanation is that the accretion we are seeing at this time is dominated by ``blobby'' accretion, and as each blob of material accretes we see an associated increase in brightness.

\subsubsection{AM Her}

\begin{figure}
\begin{center}
\includegraphics[width=\columnwidth]{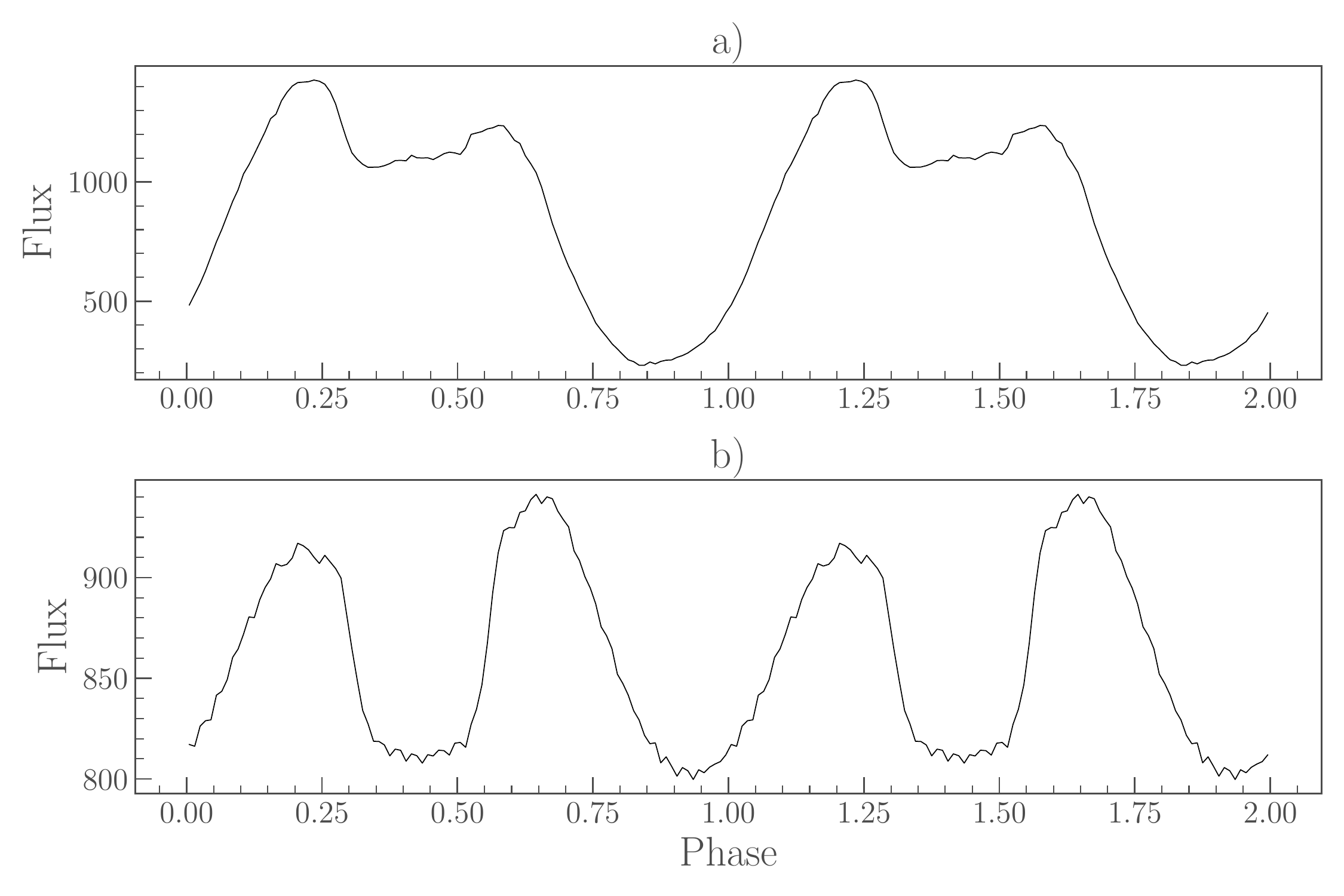}
\caption{Folded and binned lightcurve of {AM Her} a) during the initial high state ({BTJD=}2420-2432.6) and b) during the short low state ({BTJD=}2433.6-2446.6). Both have been phased with an epoch set to the start of sector 41.}
\label{fig:foldingAMHer}
\end{center}
\end{figure}
%{\textbf{The first paragraph on this section needs rewritten to actually reflect that these LCs are different. In shortening this section that meaning has been lost from the text. The second para needs revisted about the LCs being dominated by cyclotron to clear up why they don't look the same}}

In \autoref{fig:foldingAMHer} we show the folded and binned orbital lightcurve of the high state in AM Her and the short duration low state which immediately followed, both of which were seen in sector 41. The low state has a distinctly different orbital lightcurve from that seen in the high state with two clear stand-alone peaks. Although the peaks in the low state occur at approximately, but not exactly, the same phase as in the high state, albeit at much lower flux values, the relative strength of these peaks are reversed as the first peak shows a greater reduction in flux than the second.

%In \autoref{fig:foldingAMHer} we show the folded and binned orbital lightcurve of the high state in {AM Her} and the short duration low state which immediately followed it in sector 41. Both of these lightcurves are double peaked at the same phase values, however the location of the strongest peak is reversed between the two states. Furthermore, the peaks are substantially less bright in the low state; with a much smaller overall brightness variation. As such whilst the drop in brightness at between the two peaks in the low state appears to be greater than the similar drop in the high state it is proportionally smaller ($\sim13.5\%$ in low vs. $\sim23.4\%$ in high).

%at phase=0.21 and phase=0.60, however in the high state the earlier peak is the brighter of the two, whereas in the low state the converse is true.  The reason for this visual difference is that the second peak in the low state is more isolated than that in the high state, and that the total scale of the brightness variation in the high state is more than an order of magnitude greater. This is best illustrated by comparing the minimas in each lightcurve; whilst there is a substantial difference in the brightness observed between them in the high state, they carry almost identical values in the high state.

The orbital lightcurve during the high state which we have presented here is consistent with that found by \citet{2001A&A...372..557G} when observing the high state of {AM Her} in the \textit{V} band. They identified this lightcurve to be dominated by cyclotron emission from the main accretion pole. \citet{2005AJ....130.2852K} performed a multi-wavelength study of {AM Her} in a low state; and in particular obtained photometry in \textit{JHK} bands. The TESS low state observations here match the \textit{J} and \textit{H} band observations, whereas the \textit{K} band lightcurve was more reminiscent of the high state. Their findings confirmed that accretion via Roche-lobe overflow had effectively ceased, but that the profile was dominated by cyclotron emission even in this low mass-transfer state. Our findings are entirely consistent with these observations.

\subsubsection{CW Hyi}

\begin{figure}
\begin{center}
\includegraphics[width=\columnwidth]{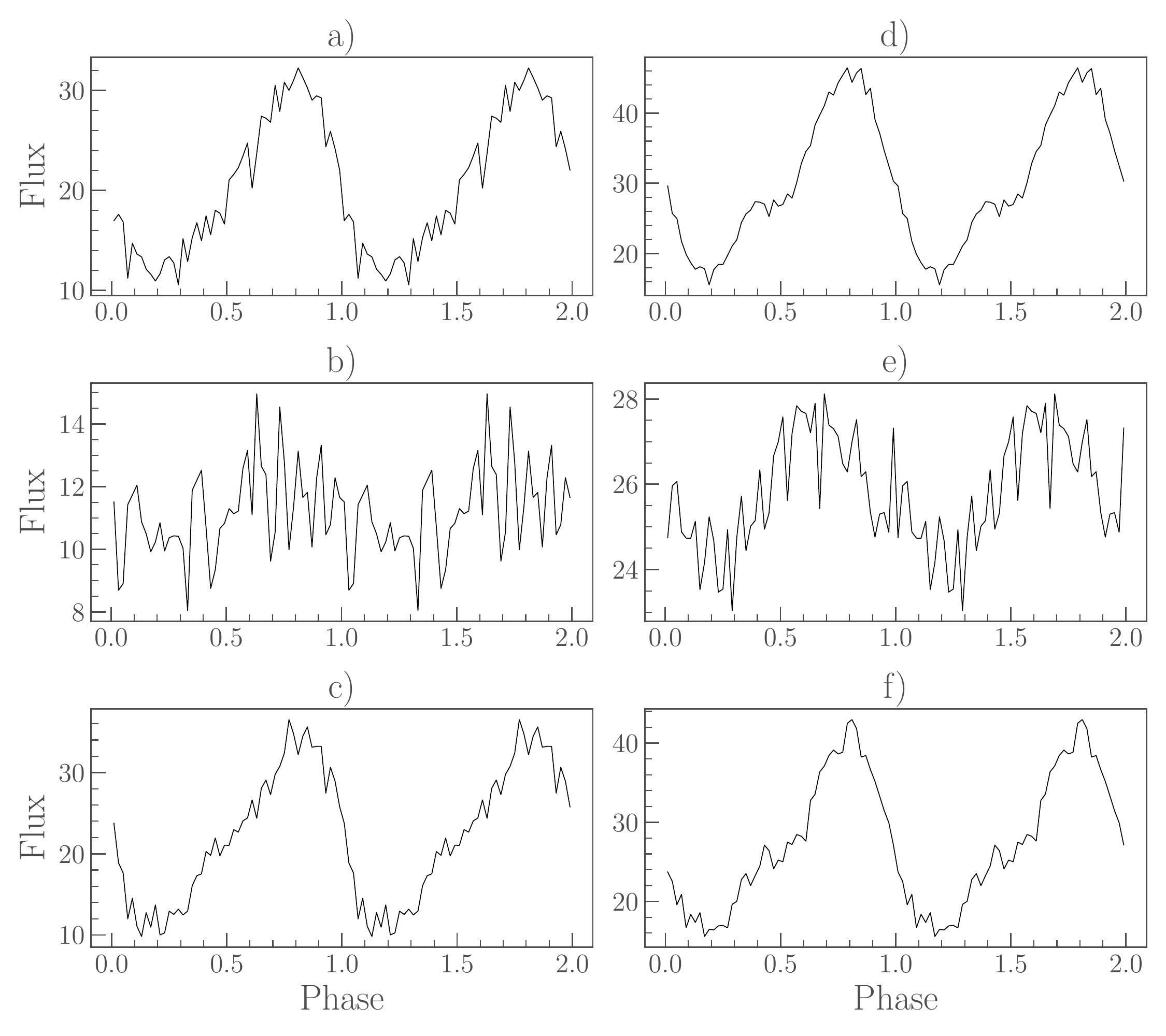}
\caption{Folded and binned lightcurve of {CW Hyi} a) before the sector 27 dimming Event ({BTJD=}2053.8-2054.9), b) during the short duration low-state Event ({BTJD=}2054.9-2055.9), c) after the Event ({BTJD=}2055.9-2056.9), d) before the sector 28 dimming Event ({BTJD=}2066.5-2068.5), e) during the short duration low-state Event ({BTJD=}2068.5-2070.5), and f) after the event ({BTJD=}2070.5-2072.5). All of the profiles have been phased with an epoch set to the start of sector 27.}
\label{fig:foldingCWHyi}
\end{center}
\end{figure}

In \autoref{fig:foldingCWHyi}, we show the orbital lightcurves of {CW Hyi} before, during and after the short reductions in brightness seen in sectors 27 and 28. We found that folding the lightcurve upon the 181.8 min period reported by \citet{2002A&A...396..895S} resulted in a phase drift of 0.3 cycles per \textit{TESS} sector. Investigation of the \textit{TESS} data revealed an orbital period of 181.55 min. This improvement in the period eliminated the phase drift and thus we have used it for the remainder of our analysis.

The orbital lightcurves before and after each event are very similar; this is in contrast to the lightcurves of the two lower states which show differences. Though it is possible to identify the original shape of the orbital lightcurve in the sector 27 event it is substantially diminished such that we believe that this marks a $\sim1.5$ d cessation in accretion. The event in sector 28 on the other hand does not have the same brightness reduction, which is likely why the shape can be more easily recovered. In this event we do not believe that accretion has ceased; only reduced whilst retaining a similar accretion geometry. We were able to access limited contemporaneous All Sky photometry via AAVSO\footnote{The American Association of Variable Star Observers: \url{aavso.org}} which indicated that outwith the events seen {CW Hyi} was in the same brightness state as it is usually found (mag $\simeq16.5-18.5$). This supports our contention that these are very short durarion, possibly unstable, low states where accretion is dramactically reduced or stopped entirely.

Comparing those orbital lightcurves outwith both of the events discussed here with those presented by \citet{2002A&A...396..895S} and \citet{2010IBVS.5957....1C} shows good agreement in both shape and brightness variation. \citeauthor{2002A&A...396..895S} identified this single peaked feature to be the signature of single pole accretion with the flux originating from beamed cyclotron emission. No published work presents orbital lightcurves of CW Hyi which appear similar to those of the dimming event nor have such events been reported before, consequently we believe that this is the first time a state transition of any sort has been identified in CW Hyi.

\subsubsection{MT Dra}

\begin{figure}
\begin{center}
\includegraphics[width=\columnwidth]{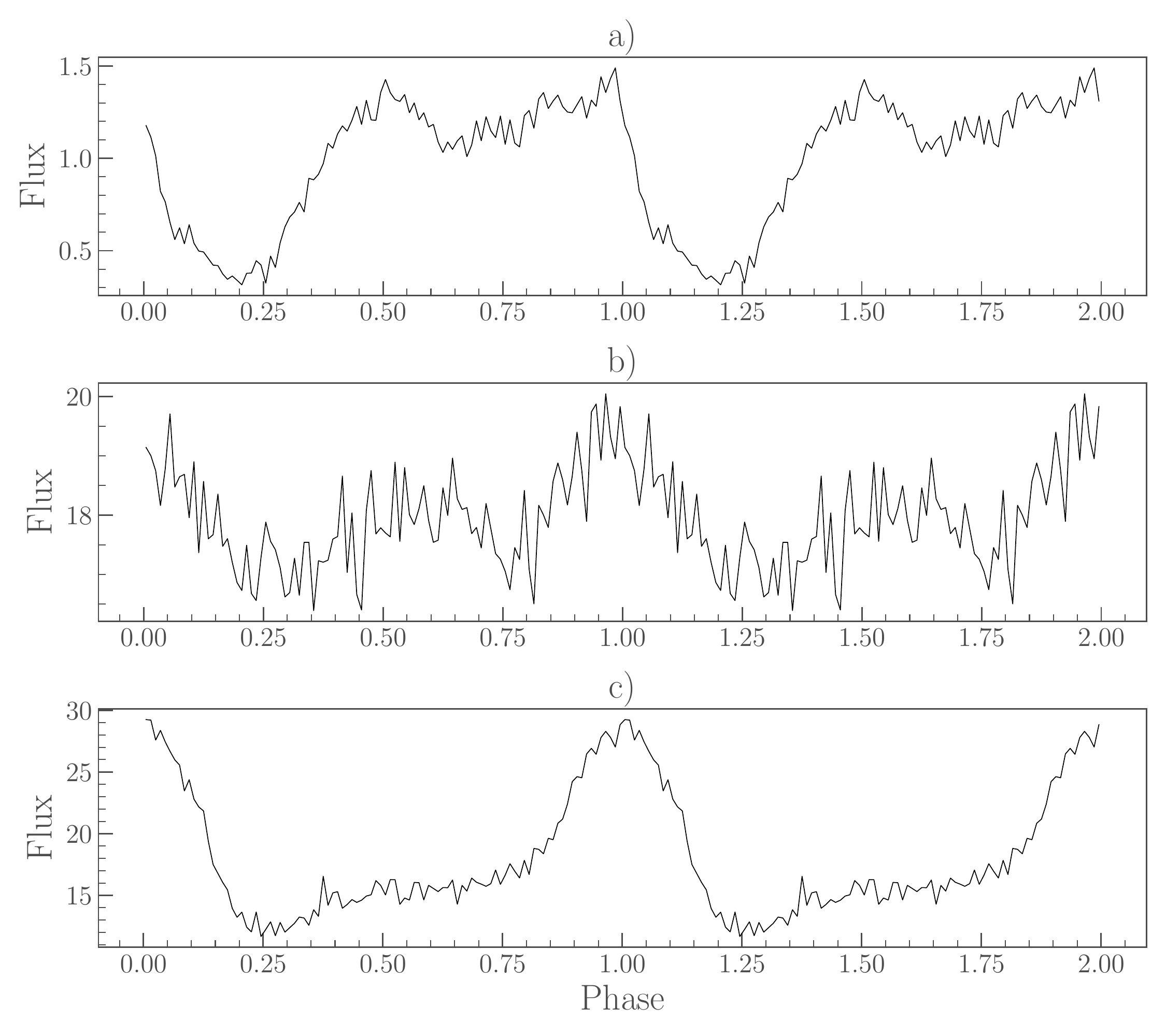}
\caption{Folded and binned lightcurve of {MT Dra} a) the high state seen during sector 26 ({BTJD=}2023.1-2035.1) b) during the Event ({BTJD=}2390.6-2404.4), c) after the Event ({BTJD=}2407.5-2418.9). All three have been phased with an epoch set to the start of sector 40.}
\label{fig:foldingMTDra}
\end{center}
\end{figure}

In \autoref{fig:foldingMTDra} we present the folded and binned lightcurves of MT Dra before, during and immediately after the low state centred on MJD=59347 (see \autoref{ZTfPlots} c)). The orbital lightcurve before and during the event both show two peaks, suggesting that two pole accretion is taking place, albeit at markedly different rates. However, without polarimetry it is not possible to confirm this, as it is possible that cyclotron beaming has produced two maxima from emission of from a single pole. As both peaks appear at the same phase, we believe that matter is accreting along the same field lines onto the same regions in each case. The single peaked nature of the state after the event suggests that this state is one of single pole accretion with matter accreting along one of the footpoints of field lines seen in the two pole case.

%The orbital lightcurves of the proceeding high state shows an initial decrease to a minima at phase=0.20 before increasing to a maximum at phase=0.50. Following this the brightness drops to a local minima at phase=0.60 before rising again until a maximum at phase=1.0. The lightcurve of the low state shows a decrease to a minimum brightness at phase=0.30; this is followed by a rise to a local maximum at phase=0.60. Succeeding the local maximum the brightness decreases to a local minimum at phase=0.80 before again rising to the global maximum at phase=1.0. The orbital lightcurve of the high state after the low state shows a somewhat different behaviour from the earlier high state with the same initial fall to a minimum brightness at phase=0.25 which is followed by a steady increase in brightness until phase=0.80 after which the the brightness steeply rises to its maximum at phase=1.0. These differences, namely the absence of a second peak in the second high state profile, suggest that the whilst both the initial high state and the low state show two pole accretion, albeit with substantially different mass transfer rates, the second high state does not.

Our findings differ from \citet{2011ARep...55..224Z} who identified a single peaked lightcurve in the low state and a double peaked lightcurve in the high state. Furthermore, whist the scale of the brightness variation across the orbital period in both high states is consistent with these earlier findings the scale of the brightness variations in the low state is markedly reduced from $F_{1}/F_{2} \simeq5.7$ in that work to $F_{1}/F_{2} \simeq0.1$. \citeauthor{2011ARep...55..224Z} reports evidence for an extended low state lasting longer than reported here, additionally being $>$1 mag brighter compared to \autoref{ZTfPlots} c) though the brightness of both high states are consistent. This implies that we have identified an even lower ``low'' state in {MT Dra} where accretion occurs onto two poles.

It is not clear why the high state which immediately follows the low state has a different profile. Although there maybe several reasons for this, we note that the flow through L1 is affected by magneto-hydrodynamic (MHD) conditions at L1. The local magnetic field configuration depends on the combination of local spot fields, the global donor field, and the WD magnetic field. For the light curve to return to the same shape, the local field configuration must also return to a comparable state as it directly affects mass transfer through the L1 region.

\section{Discussion}

Low states in Polars are a natural consequence of a substantial decrease in the mass-transfer rate or a temporary disruption of the mass-transfer process. While the former can be caused by the secular evolution of the binary, both the former and the latter can be caused by an alternation in the MHD properties of the companion star near the L1 point. A possible cause of the suppression of mass outflow from the companion star is the adjustment of the topology of the magnetic field at the L1 region, which may be a response to the change in the magnetic configuration set by the WD and the companion star \citep[see][]{2008A&A...481..433W}. Another possibility is the magnetic field near the L1 point is readjusted in response to the migrating in of a large star spot or a cluster of star spots \citep[see][]{1994ApJ...427..956L,2016MNRAS.463.1342S}. The essence of these scenarii is that the magnetic field configuration at the L1 region is modified. It does not require an atmospheric response of the whole companion star that would in turn invoke more complex mechanisms, such as irradiation by the accreting WD, which leads to a building up period for the mass-transfer and irradiation feedback \citep[see e.g.][]{Wu1995PASA}. This scenario in the framework of magnetic star spots is therefore more appealing and is of particular relevance in the context of the transitions into low states which we present here.

VY Scl stars are a subclass of nova-like (NL) CVs which exhibit pronounced low states; they are a relatively rare subclass within the NL CVs with \citet{2004AJ....128.1279H} noting that from a sample of 65 NLs only $\sim15\%$ of them exhibited the phenomenon and of that sample only a single system exhibited a state change of similar duration to those presented here. Unlike Polars however, VY Scl systems have accretion discs and lie almost exclusively above the period gap. It is thought that VY Scl stars move between the two stable configurations for disc accretion offered by the disc accretion model; a hot large disc for the high state and a small cool disc for the low state \citep{warner_1995}. The long running challenge with these systems has been to explain why they do not show outbursts during the transition between states; with competing models suggesting a truncated disc cause by irradiation of the inner disc \citep{1999MNRAS.305..225L} or considering that the WD is magnetic and that VW Scl stars are in actual fact IPs \citep{2002A&A...394..231H}. Despite lacking a comprehensive understanding of the states in these systems, it is still instructive to compare the low states with those we have explored here.

It is clear that although similarities exist between low states in Polars and those seen in other types of CVs; there are a number of significant differences. The observations which we present here indicate that Polars are capable of short duration low states and exist in a number of systems such as to make it a not uncommon occurrence, whereas in IPs and VY Scl systems this is uncommon.

In the following discussion we will explore the short duration low and high states and their possible physical origins and how these vary, or otherwise, with the states seen in over systems. Furthermore we explore the changes in the orbital lightcurves as a result in these changing brightness states and how that informs the accretion geometry of these systems.

\subsection{Short-lived Low States}
%{\textbf{Need to bring in the models that already exist in a bit more detail and how these might be explained etc. Probably something for Kinwah to build out a little based on our conversations}}

Out of those systems which we identify as containing a short-lived state change the most common manifestation of this was a low state. In 11 cases, 9 in the ZTF sample and 2 in the TESS sample, we observed multiple events pointing to the origin of these short duration states being a feature that is relatively common within Polars that can form with ease as opposed to requiring rigorous or difficult to establish conditions. Despite having determined recurrence times for a number of systems we found no correlation between this and either $B$-field strength of the WD or orbital period; similarly, no such relation could be robustly developed with respect to the duration of these states. This notwithstanding, however, we do note that the shortest duration events seen only in TESS appear to originate in systems with longer periods; though the small number of systems that showed these shortest duration states mean that this may be a statistical quirk. Additionally, though no obvious correlations with orbital period appear to exist we do note that the systems which we see to posses short-lived state changes appear to be predisposed to lie below the period gap at a rate greater than the overall population of Polars.

The duration of these short-lived low states appears to be variable within our definition of short-lived; between a few days up to $\sim50$ d. The duration of these low-states and their common occurrence essentially rules out that low-states are related to thermo-hydrodynamic activities of the companion. This can be seen as follows. The thermodynamic adjustment of star in response to external perturbation is constrained by the Kevin-Helmholtz time scale, which may be expressed in terms of the stellar mass $M$, stellar radius $R$ and stellar luminosity: 
\begin{equation} 
\tau_{\rm kh}  = \frac{GM^2 \xi}{2 RL} \approx 
 3.1 \times 10^7\;\! {\rm yr} 
  \left(\frac{M}{{\rm M}_\odot} \right)
  \left(\frac{R}{{\rm R}_\odot} \right)^{-1}
  \left(\frac{L}{{\rm L}_\odot} \right)^{-1}\;\! \xi  \  , 
\end{equation}
where $G$ is the gravitational constant. Here, we introduce a variable $\xi = \Delta M/M$, where $\Delta M$ is mass of the layer at the stellar surface participating in the thermo-hydrodynamic response. Setting $\xi = 1$ will recover the standard expression for the Kevin-Helmholtz timescale of star, which is commonly used for stellar response to mass loss \citep[see e.g.][]{Plavec1973PASP}. 
For low-mass stars with $\left( M/{\rm M}_\odot\right) \sim (0.2 - 0.4)$, $\left({M}/{{\rm M}_\odot} \right) \sim\left({R}/{{\rm R}_\odot} \right)$ \citep[see e.g.][]{Demory2009A&A}. Also, $\left({L}/{{\rm L}_\odot} \right) \sim\left({M}/{{\rm M}_\odot} \right)^4$ \citep[see e.g.][]{Wang2018A&A}. This gives $\tau_{\rm kh} \sim 3 \times 10^7 {\rm yr} \;\!
\left({M}/{{\rm M}_\odot} \right)^{-4}\;\! \xi \gg 1~{\rm yr}$ for $(M/{\rm M}_\odot)\sim 0.2$, even when assuming an extremely small mass layer with $\xi \sim 10^{-7}$ \citep[cf.][]{Kovetz1988ApJ,Sarna1990A&A,Ginzburg2021MNRAS}. Thus, the thermo-hydrodynamic response of the companion is unlikely to be the cause of the short-lived low states observed in Polars.

The magnetic activity of the system is, however, a viable alternative. It involves changes in the magnetic field configuration of the binary, which in turn disrupts the mass transfer process. This can be achieved, for instance, when a star spot on the companion's surface migrates to a region near the L1 point \citep[see][]{1994ApJ...427..956L}. Rotating convective stars are magnetically active, and star spots are expected to be present in the companion stars in CVs, including both Polars and IPs. In Polars, the magnetic field of the WD can also play a role as it is strong enough to influence the magnetic field configurations of the entire system. 
The magnetic field near the L1 point is not always dominated by the local magnetic field of the companion star, as the magnetic field of the WD and that of the companion star may be interconnected. As such, the mass transfer process is also associated with the interaction of the magnetic fields of the immigrated star spot and of the WD at the L1 point \citep[see e.g.][for the influence of the WD magnetic fields on the mass transfer]{2008A&A...481..433W}. While this coordinated magnetic interaction would lead to the short low states by the suppression of the mass transfer, it can also lead to brief mass transfer episodes fulled by (active) magnetic channelling, which manifests as the short high stated seen in some Polars. The short-lived low states and the short-lived high states are therefore facets of the magnetic-interaction regulated mass transfer dynamics, and we shall discuss this more thoroughly in the next subsection, in the context of mass transfer models, \citep[e.g.][]{Lubow1975ApJ,Meyer1983A&A,Osaki1985A&A,Ritter1988A&A}. 

\subsection{Short-lived High States}
%{\textbf{These seem relatively novel and limited only to Polars -- though you would note that in other CVs you'd expect to see outbursts as the short (unstable) enhanced accretion state. But I am not entirely sure what to say beyond that just yet. I have spoken with Boris today about these systems, suggested that it's a break in magnetic breaking that briefly increases the mass transfer rate (aware I have phrased this poorly, just adding while it's still at the front of my mind.}}

%{\textbf{Please see the short-lived high state in Figure \ref{fig:} - PM}}

Short-lived high states are by far the least common feature which we have identified in this work -- having identified only 3 systems. Despite this they show clear evidence of being repeated features with {AP CrB} showing evidence for as many as five events over $\sim1400$ d. These events are a result of a temporary increase in the rate of mass transfer within the system.

Much of the population of CVs which as IPs or Dwarf Novae are known to be capable outbursts, even if individual systems have not been observed to do so. These outbursts are relatively short duration (tens of days) events whereby accretion, by virtue of an instability, suddenly and temporarily increases. These outbursts are characterised broadly via their asymmetric temporal evolution; a sudden increase in brightness over a day or less, followed by a short plateau phase, and finally an gradual reduction in brightness \citep[see][for a fuller discussion of CV outbursts]{2016MNRAS.456.4441C}.

Conversely, Polars are not known to be systems capable of dwarf nova outburst; because they lack the necessary accretion disc. Similarly the short duration high states, which we report in this work, do not have the same temporal profile as outbursts seen in other systems -- they are more symmetric in shape, resembling transitions in and out of previously observed long duration states. As such we are not proposing that these events are outbursts in Polars, but the very short duration does suggest that these state changes are unstable.

Accepting this, and further that the short duration low states are related to star spots, we can explain the occurrence of the short-lived high states in a self-consistent manner, i.e. within the same framework of the coronal/atmospheric magnetic activities of the companion star. Doing so requires that we address several issues. 

Firstly, we must consider whether or not these high states are consequences of thermo-hydrodynamics activities. As discussed previously, this will require a timescale comparable to the Kevin-Helmholtz timescale. Applying the same argument as before, it is unlikely that thermo-hydrodynamics drives the system into such short episodes of high state. 
  
Next, we need to look at whether or not there are sharp changes in the physical condition near the L1 point, where the mass transfer initiates. There are certain conditions that need to be satisfied for mass transfer to occur when the companion star is reaching the critical Roche surface of the binary. One of which is that the outflow must become transonic when the material crosses the L1 point. This gives a simple approximate expression for the mass transfer rate: 
\begin{equation} 
  {\dot M} \sim \left[ \rho \;\! c_{\rm s}(T)\right]_{\rm L1}\;\! {\cal A}   
 \end{equation} 
\citep[see e.g.][]{Lubow1975ApJ,Meyer1983A&A}, where the mass density $\rho$ and the isothermal sound speed $c_{\rm s}$ and the stellar surface temperature $T$ are evaluated at the L1 point at the critical Roche surface and ${\cal A}$ is the initial cross section of the mass-outflow stream. As a first approximation, the atmospheric mass density at the L1 point would take the form 
 \begin{equation} 
  \rho \approx \rho_0   \ 
    \exp\left[- \frac{1}{2}\left(\frac{\Delta r}{z_0} \right)^\alpha \right]  \ ,    
\end{equation}  
where $\rho_0$ is density at the ``base'' of the stellar atmosphere\footnote{{{The star is not spherical and the atmospheric base is ill defined. However, a reference location for the purpose of this work may be defined in terms of the optical depth, and following \cite{Kovetz1988ApJ}, we may take it as the location in the photosphere with optical depth, $\tau$,  
at 2/3.}}}, 
$\Delta r$ is the separation length between the critical Roche surface and the surface of the ``base'' of the stellar atmosphere, $z_{0}$ can be considered as a reference scale height of the stellar atmosphere\footnote{Usually, for an isothermal stellar atmosphere, the atmospheric scale is given by $z_{0} \sim k_{\rm B}\;\! T/\mu m_{\rm H}\;\! g$, where $k_{\rm B}$ is the Boltzmann constant, $m_{\rm H}$ is the hydrogen mass, $\mu$ is the mean molecular weight, $T$ is the atmospheric temperature, and $g$ is the surface gravity.}. The index $\alpha$ may take a value between 2 for semi-detached binaries, with consideration of the forces associated with binary rotation and atmospheric hydrodynamic in the companion star \citep{Lubow1975ApJ,Meyer1983A&A,Kovetz1988ApJ}, and 1, for a hydrostatic atmosphere \citep{Osaki1985A&A,Ritter1988A&A}.  
  
%The dependence on $(\Delta r)$ or $(\Delta r)^2$ in the mass transfer rate above can be %derived from the consideration of Roche equi-potential surfaces $\phi (r)$, i.e. $\dot M %\sim \exp \left[- (\phi - \phi_{0}) \right]$, where $\phi_0$ is the Roche potential at the %base of the stellar atmosphere $r_0$. From the Taylor expansion of Roche potential 
%\begin{equation}  
%\phi - \phi_0 = \frac{\partial \phi}{\partial r}\bigg\vert_{\rm r_0} (\Delta r) 
%  +  \frac{\partial^2 \phi}{\partial^2 r}\bigg\vert_{\rm r_0} (\Delta r)^2 
%    \ \cdots   
%    \ . 
%% \nonumber \\  
% & \approx \frac{\partial \phi}{\partial r}\bigg\vert_{\rm L1} (\Delta r) 
%  +  \frac{\partial^2 \phi}{\partial^2 r}\bigg\vert_{\rm L1} (\Delta r)^2 
%    \ \cdots   \ , 
%\end{equation} 
%As L1 is a saddle point, if we set 
%${\partial \phi}/{\partial r}\vert_{\rm r_0}= {\partial \phi}/{\partial r}\vert_{\rm L1}$, %the leading term of the expansion would be proportional to $(\Delta r)^2$, as shown in %\cite{Lubow1975ApJ}. 

In the presence of a magnetic field, the stellar material flowing from the secondary star will be subject to both gravitational and electromagnetic forces. Thus, the criterion in terms of solely the Roche potential is insufficient, as one needs to account for the electromagnetic force. We may decompose the force density exerted on the stellar material of density $\rho$ near L1 point as 
 \begin{equation} 
  {\boldsymbol f} = {\boldsymbol f}_{\rm g} +{\boldsymbol f}_{\rm B}  
   = -\rho\;\!  \nabla \;\! \phi_{\rm g} +{\boldsymbol f}_{\rm B}  \ , 
\end{equation}
with the subscripts ``g'' and ``B'' denoting gravitational origin and electromagnetic origin respectively, and $\phi_{\rm g}$ is the effective gravitational potential. The magnetic force density acting on the (partially) ionised material is given by  
\begin{equation}  
 {\boldsymbol f}_{\rm B} = 
  \frac{1}{4\pi} 
   \left[ \left({\boldsymbol B}\cdot \nabla \right)\;\! {\boldsymbol B} 
   - \frac{1}{2}\;\! \nabla\;\! B^2\right]  \ ,   
\end{equation} 
where the first and second terms in the bracket correspond to the magnetic tensor force and the magnetic pressure gradient force respectively. The strength of these forces on the stellar material depend on the degree of ionisation of the stellar material near the L1 point, in addition to the magnetic-field strength and orientation. It is important to note that these magnetic forces are local quantities near the L1 point, contributed mostly by the presence of an migrating star spot which interacts with the WD magnetic field. This is in contrast to $\phi_{\rm g}$, which is a global quantity of the binary defined by the gravitational field of the WD and the companion star and their orbital motion. 

The presence of a local $\phi_{\rm B}$ implies that the mass transfer rate is no longer regulated solely by the gravitational force within the binary and the thermo-hydrodynamic condition of the companion. Variations in $\phi_{\rm B}$ near the L1 point would lead to variations in $\dot M$ regardless of other conditions in the companion star. We have attributed the low states as the disruption  of on-going mass transfer by an in-migration of a star spot, such that a magnetic force temporally suppresses the outflow of material through the L1 point. In the same token, we may consider that the occurrence of the short high states corresponds to a temporary enhancement of mass flow across the gravitational surface through the magnetic channels. Such an enhancement may arise via interconnecting the WD magnetic field, when the companion star is very close to, but has not yet reached, the critical state that mass transfer proceeds in a natural manner \citep[cf.][]{Bonnet-Bidaud2000A&A}.

Flow suppression and enhanced magnetic channelling are caused by the migration of the star spot to regions close to the L1 point, and so they would operate on similar timescales, at least in the first order approximation. Moreover, the high and low state lightcurves should share some inverted up-down similarity, which we see in the lightcurves presented here. This extension of the magnetic interaction mechanism such as that outlined in \cite{2008A&A...481..433W} to incorporate the in-migration of star spots could explain the presence of short low states and short high states, and the existence of both short low states and short high states of the same system at different observational episodes, though we have seen no evidence of such systems. It overcomes some difficulties in the earlier versions of the star spot model, e.g. the requirement that the high state is the default mass transfer state \citep[see][]{1994ApJ...427..956L} and explains short high and low states being more frequently found in Polars rather than in other CVs.

{{We now remark on why the magnetic field could still confine material outflow from the donor stars in Polars, which are generally low-mass M stars, with optical/IR photospheric emission with spectral temperatures of $2000-3000~{\rm K}$. We expect that gas at $2000-3000~{\rm K}$ is not ionised and thus as neutral gas would not be magnetically confined when the Lorentz force is absent. Magnetic confinement can occur due to several very specific conditions in these systems. First of all, these low-mass M stars have magnetic coronal activities which produce high-energy charged particles. These charged particles are confined by the magnetic field and the neutral material is confined through the collision with the confined charged particles; the neutral material can only cross the magnetic field through ambipolar diffusion \citep[see e.g.][]{Spitzer1978}. This together with the transonic condition for material outflow from the L1 point \cite[see][]{Lubow1975ApJ} gives the confinement condition $\chi  n_{\rm p}  > 10^3{\rm cm}^{-3}$, where $\chi$ is the ionisation fraction and $n_{\rm p}$ is the number density of particles at the L1 point (see the derivation in Appendix \ref{appx:ambi}). Provided that a non-negligible amount of ions are present and the material near the L1 point is sufficiently dense (so to maintain strong collision between particles), the magnetic field will have strong effect outflow of cool neutral gas. 
}}

\subsection{Mass transfer rate during the transition between high and low states} 

The magnetic scenario linking the migration of star spot(s) to the L1 point and the interaction between the magnetic fields of the star spot and the WD seems to provide an adequate explanation of both the short-lived low states and the short-lived high states, even although the mechanisms likely differ in detail.

The successive occurrence of these states implies that clusters of star spots, instead of a giant isolated star spot, is likely responsible. A remaining issue is what observational signature(s) would we expect for the mass transfer rate if these short state transitions are caused by magnetic activities near the L1 point when a high state transitions to a low state or vice versa. This magnetic interaction scenario does not require substantial changes in the global thermal structure of the companion star, and hence the variations and the stability in mass transfer are not driven by the variations in thermodynamics variables of the outer layer of the companion star. Instead, they are regulated by the magnetic field configurations, which are determined by the migrations of the star spot, and the relative strength and orientation of the magnetic fields of the star spots and of the WD. 
  
The magnetic field configuration at the L1 point can also be adjusted through processes, such as MHD interactions between the ionised material and the magnetic field at the L1 point and also events such as magnetic reconnections, which are eruptive in nature. %see Figure \ref{}, where first there is a reduction in the mass transfer rate, consistent with a star-spot crossing at L1, during which a sudden outburst occurred.
As such, we would expect that the mass transfer rate would not change smoothly, when a Polar switches from a low state to a high state or from a high state to a low state. Flare like activities are expected if magnetic reconnections occur, as the mass outflow will be broken up into magnetised fluid chunks \citep[see for instance the flare on the Polar MQ Dra][]{Ramsay2021}. This is also in contrast to the abrupt yet smooth change in the source brightness when it is caused by occultation \citep[cf. e.g. the eclipse in the photometric light curves of  CTCV~J192$-$5001][]{Potter2005MNRAS}. 

\subsection{Orbital folded light curves and mass transfer rate}

%{\textbf{Going to be adding some text with thoughts on what the changes in the orbital lightcurves potentially tell us about the changes in the accretion region of the surface of the WD, potentially will be hard to do with any great accuracy with the data that we have but we might be able to point to a direction of travel on this topic. \\ \\Currently think that this is too long talking about the systems that we have seen in a recap. Will need to cut down}}\\

The TESS data has allowed us to search for differences in the profile of the optical orbital lightcurves as a function of mass transfer rate (i.e. high/low states).  In turn, this allowed us to address whether the stream couples onto different field lines depending on the accretion rate and therefore its ram pressure. At high mass transfer rates the ram pressure is high and can penetrate deeper into the WD magnetic field. This allows for material to reach field lines which lead to a second accreting pole on the WD which under a lower mass transfer rate, and consequently ram pressure, they might not. The alternative model of ``blobby'' accretion was proposed by \citet{1982A&A...114L...4K} whereby individual ``blobs'' of material penetrate the WD photosphere; this model has been used to explain the apparent difference in behaviour seen from the different poles \citep{1988MNRAS.235..433H,2020A&A...642A.134S} and the soft X-ray excess seen in many Polars \citep{FrankKingLasota1988}.

A further possible origin for the change of accretion geometry is asynchronous rotation whereby the WD has a (temporarily) marginally different spin period from the orbit of the system. The material latches onto different field lines as a result of the differential rotation; meaning the material can penetrate to different depths depending on the field lines in use. This change however, is not associated with a change in mass transfer rate and thus is unlikely to apply in any of the systems presented in this work. 

In each of those systems in this study where we see a low state, we also observe an associated change in the shape of the lightcurve; the nature of which seems somewhat variable, e.g., the first low state seen in {CW Hyi} shows a cessation of accretion, whereas the second shows it to continue in a similar way to as before the state but at a much suppressed rate. In {AM Her} and {QQ Vul} we see the increased pronouncement of a second peak in the profile which has been observed before and are considered to be the revealing of features present during the high state but otherwise dominated by the primary accretion stream. In each of these cases, where accretion continues in the low state, the original lightcurve is still recognisable within the low state. It is either supplemented by an additional feature or continued at a lower flux rate. This suggests that no significant change in accretion geometry occurs in these low states, beyond the rate of accretion falling substantially.

{MT Dra} is the markedly different system out of the four systems with short duration low states shown here; the orbital lightcurve of the low state follows a similar but suppressed profile of the brighter state following it, whereas the bright state proceeding it shows a different profile, albeit one which is related, with the same initial dip. We note that {MT Dra} is the only one of the four systems which has a period below the period gap. In contrast to those other systems {MT Dra} does appear to show a changing accretion geometry between the states.

{V834 Cen} was the only system which showed a short-lived high state which we were able to construct an orbital lightcurve. In this case although the increased brightness implied an increased accretion rate we show no evidence of a changed accretion geometry. It is therefore likely the case that the change ram pressure in the high state is either insufficient to push deeper into the WD magnetic field or that in the initial state the accretion flow already penetrates to such a depth that any increase in the ram pressure is negligible relative to the magnetic energy density at that depth.

\section{Conclusions}

We have presented observations from ZTF and TESS of short duration state changes as seen in a number of Polars. These reveal that short duration state changes are a common feature in Polars. The observations from TESS in particular have revealed a number of especially short duration state changes which without the coverage and cadence provided by TESS would have been hitherto hidden from us. The observations from TESS have allowed us to probe the accretion geometry of these systems by studying the changes in the orbital lightcurves during the various accretion states revealing how the material in-fall is altered in these different scenarios. {{The TESS observations also have allowed us to refine the orbital period of CW Hyi where we found a period of 181.55 min.}}

Considering these state changes has allowed us to generalise the scenario of star spot migration traditionally adopted for explaining the occurrence of the low states to incorporate the interaction between the magnetic fields of the star spots and that of the WD. As such we have developed a satisfactory explanation for both the short duration low and high states within a self-consistent and unified framework; whereby the mass transfer rate is consequentially determined by the thermal and magneto-hydrodynamic conditions of the companion star at the L1 point, in the presence of the magnetic interaction between the star spots and the WD.

In order to advance this work into these different state changes further useful data to gain would include polarimetry and X-ray observations as this would allow for mapping of the accretion accretion regions and understanding the mass transfer rates. In addition to this further inclusion of Polars in upcoming TESS sectors would be useful to build up a more complete understanding of the number of systems which exhibit this behaviour; in particularly the shortest duration states, the frequency of this behaviour; and its breadth across the population of Polars.

\section*{Acknowledgements}

{{We thank the referee for comments and suggestions that help to make clear some of the phenomenological modelling in this work.}}

This paper includes data collected by the TESS mission. Funding for the TESS mission is provided by the NASA's Science Mission Directorate.
It also includes ZTF data  obtained with the Samuel Oschin Telescope 48-inch and the 60-inch Telescope at the Palomar Observatory as part of the Zwicky Transient Facility project. ZTF is supported by the National Science Foundation under Grants No. AST-1440341 and AST-2034437 and a collaboration including current partners Caltech, IPAC, the Weizmann Institute for Science, the Oskar Klein Center at Stockholm University, the University of Maryland, Deutsches Elektronen-Synchrotron and Humboldt University, the TANGO Consortium of Taiwan, the University of Wisconsin at Milwaukee, Trinity College Dublin, Lawrence Livermore National Laboratories, IN2P3, University of Warwick, Ruhr University Bochum, Northwestern University and former partners the University of Washington, Los Alamos National Laboratories, and Lawrence Berkeley National Laboratories. Operations are conducted by COO, IPAC, and UW.

This work was funded by UKRI grant (ST/T505936/1). For the purpose of open access, the authors have applied a creative commons attribution (CC BY) licence to any author accepted manuscript version arising.
C. Duffy acknowledges STFC for the receipt of a postgraduate studentship.
Armagh Observatory \& Planetarium is core funded by the Northern Ireland Executive through the Department for Communities. P. Mason is funded by Picture Rocks Observatory.

This research made use of Lightkurve, a Python package for Kepler and TESS data analysis.

We acknowledge with thanks the variable star observations from the AAVSO International Database contributed by observers worldwide and used in this research.

\section*{Data Availability}
TESS data is available from the Mikulski Archive for Space Telescopes (MAST), which can be accessed at \url{https://mast.stsci.edu/portal/Mashup/Clients/Mast/Portal.html}.

ZTF data is available via the NASA/IPAC Infrared Science Archive \url{https://irsa.ipac.caltech.edu/Missions/ztf.html}


\bibliographystyle{mnras}
\bibliography{references} % if your bibtex file is called example.bib

\begin{thebibliography}{}
\makeatletter
\relax
\def\mn@urlcharsother{\let\do\@makeother \do\$\do\&\do\#\do\^\do\_\do\%\do\~}
\def\mn@doi{\begingroup\mn@urlcharsother \@ifnextchar [ {\mn@doi@}
  {\mn@doi@[]}}
\def\mn@doi@[#1]#2{\def\@tempa{#1}\ifx\@tempa\@empty \href
  {http://dx.doi.org/#2} {doi:#2}\else \href {http://dx.doi.org/#2} {#1}\fi
  \endgroup}
\def\mn@eprint#1#2{\mn@eprint@#1:#2::\@nil}
\def\mn@eprint@arXiv#1{\href {http://arxiv.org/abs/#1} {{\tt arXiv:#1}}}
\def\mn@eprint@dblp#1{\href {http://dblp.uni-trier.de/rec/bibtex/#1.xml}
  {dblp:#1}}
\def\mn@eprint@#1:#2:#3:#4\@nil{\def\@tempa {#1}\def\@tempb {#2}\def\@tempc
  {#3}\ifx \@tempc \@empty \let \@tempc \@tempb \let \@tempb \@tempa \fi \ifx
  \@tempb \@empty \def\@tempb {arXiv}\fi \@ifundefined
  {mn@eprint@\@tempb}{\@tempb:\@tempc}{\expandafter \expandafter \csname
  mn@eprint@\@tempb\endcsname \expandafter{\@tempc}}}

\bibitem[\protect\citeauthoryear{{Agrawal}, {Riegler}  \& {Rao}}{{Agrawal}
  et~al.}{1983}]{1983Natur.301..318A}
{Agrawal} P.~C.,  {Riegler} G.~R.,   {Rao} A.~R.,  1983, \mn@doi [\nat]
  {10.1038/301318a0}, \href
  {https://ui.adsabs.harvard.edu/abs/1983Natur.301..318A} {301, 318}

\bibitem[\protect\citeauthoryear{{Belle}, {Howell}  \& {Mills}}{{Belle}
  et~al.}{2000}]{2000PASP..112..343B}
{Belle} K.~E.,  {Howell} S.~B.,   {Mills} A.,  2000, \mn@doi [\pasp]
  {10.1086/316536}, \href
  {https://ui.adsabs.harvard.edu/abs/2000PASP..112..343B} {112, 343}

\bibitem[\protect\citeauthoryear{{Bellm} et~al.,}{{Bellm}
  et~al.}{2019}]{2019PASP..131a8002B}
{Bellm} E.~C.,  et~al., 2019, \mn@doi [\pasp] {10.1088/1538-3873/aaecbe}, \href
  {https://ui.adsabs.harvard.edu/abs/2019PASP..131a8002B} {131, 018002}

\bibitem[\protect\citeauthoryear{{Beuermann}, {Euchner}, {Reinsch}, {Jordan}
  \& {G{\"a}nsicke}}{{Beuermann} et~al.}{2007}]{2007A&A...463..647B}
{Beuermann} K.,  {Euchner} F.,  {Reinsch} K.,  {Jordan} S.,   {G{\"a}nsicke}
  B.~T.,  2007, \mn@doi [\aap] {10.1051/0004-6361:20066332}, \href
  {https://ui.adsabs.harvard.edu/abs/2007A&A...463..647B} {463, 647}

\bibitem[\protect\citeauthoryear{{Bonnet-Bidaud} et~al.,}{{Bonnet-Bidaud}
  et~al.}{2000}]{Bonnet-Bidaud2000A&A}
{Bonnet-Bidaud} J.~M.,  et~al., 2000, \aap, \href
  {https://ui.adsabs.harvard.edu/abs/2000A&A...354.1003B} {354, 1003}

\bibitem[\protect\citeauthoryear{Brainerd \& Lamb}{Brainerd \&
  Lamb}{1985}]{BrainerdLamb}
Brainerd J.~J.,  Lamb D.~Q.,  1985, in Lamb D.~Q.,  Patterson J.,  eds,
  Cataclysmic Variables and Low-Mass X-Ray Binaries. Springer Netherlands,
  Dordrecht, pp 247--256

\bibitem[\protect\citeauthoryear{{Burwitz}, {Reinsch}, {Beuermann}  \&
  {Thomas}}{{Burwitz} et~al.}{1997}]{1997A&A...327..183B}
{Burwitz} V.,  {Reinsch} K.,  {Beuermann} K.,   {Thomas} H.~C.,  1997, \aap,
  \href {https://ui.adsabs.harvard.edu/abs/1997A&A...327..183B} {327, 183}

\bibitem[\protect\citeauthoryear{{Cieslinski}, {Rodrigues}, {Silva}  \&
  {Diaz}}{{Cieslinski} et~al.}{2010}]{2010IBVS.5957....1C}
{Cieslinski} D.,  {Rodrigues} C.~V.,  {Silva} K.~M.~G.,   {Diaz} M.~P.,  2010,
  Information Bulletin on Variable Stars, \href
  {https://ui.adsabs.harvard.edu/abs/2010IBVS.5957....1C} {5957, 1}

\bibitem[\protect\citeauthoryear{{Coppejans}, {K{\"o}rding}, {Knigge},
  {Pretorius}, {Woudt}, {Groot}, {Van Eck}  \& {Drake}}{{Coppejans}
  et~al.}{2016}]{2016MNRAS.456.4441C}
{Coppejans} D.~L.,  {K{\"o}rding} E.~G.,  {Knigge} C.,  {Pretorius} M.~L.,
  {Woudt} P.~A.,  {Groot} P.~J.,  {Van Eck} C.~L.,   {Drake} A.~J.,  2016,
  \mn@doi [\mnras] {10.1093/mnras/stv2921}, \href
  {https://ui.adsabs.harvard.edu/abs/2016MNRAS.456.4441C} {456, 4441}

\bibitem[\protect\citeauthoryear{{Covington} et~al.,}{{Covington}
  et~al.}{2022}]{2022ApJ...928..164C}
{Covington} A.~E.,  et~al., 2022, \mn@doi [\apj] {10.3847/1538-4357/ac5682},
  \href {https://ui.adsabs.harvard.edu/abs/2022ApJ...928..164C} {928, 164}

\bibitem[\protect\citeauthoryear{{Cropper}}{{Cropper}}{1990}]{1990SSRv...54..195C}
{Cropper} M.,  1990, \mn@doi [\ssr] {10.1007/BF00177799}, \href
  {https://ui.adsabs.harvard.edu/abs/1990SSRv...54..195C} {54, 195}

\bibitem[\protect\citeauthoryear{{Cropper}}{{Cropper}}{1998}]{1998MNRAS.295..353C}
{Cropper} M.,  1998, \mn@doi [\mnras] {10.1046/j.1365-8711.1998.01333.x}, \href
  {https://ui.adsabs.harvard.edu/abs/1998MNRAS.295..353C} {295, 353}

\bibitem[\protect\citeauthoryear{{Cropper}, {Menzies}  \& {Tapia}}{{Cropper}
  et~al.}{1986}]{1986MNRAS.218..201C}
{Cropper} M.,  {Menzies} J.~W.,   {Tapia} S.,  1986, \mn@doi [\mnras]
  {10.1093/mnras/218.2.201}, \href
  {https://ui.adsabs.harvard.edu/abs/1986MNRAS.218..201C} {218, 201}

\bibitem[\protect\citeauthoryear{{Demory} et~al.,}{{Demory}
  et~al.}{2009}]{Demory2009A&A}
{Demory} B.~O.,  et~al., 2009, \mn@doi [\aap] {10.1051/0004-6361/200911976},
  \href {https://ui.adsabs.harvard.edu/abs/2009A&A...505..205D} {505, 205}

\bibitem[\protect\citeauthoryear{{Ferrario}, {Wickramasinghe}, {Bailey},
  {Tuohy}  \& {Hough}}{{Ferrario} et~al.}{1989}]{1989ApJ...337..832F}
{Ferrario} L.,  {Wickramasinghe} D.~T.,  {Bailey} J.,  {Tuohy} I.~R.,   {Hough}
  J.~H.,  1989, \mn@doi [\apj] {10.1086/167153}, \href
  {https://ui.adsabs.harvard.edu/abs/1989ApJ...337..832F} {337, 832}

\bibitem[\protect\citeauthoryear{{Foreman-Mackey}, {Hogg}, {Lang}  \&
  {Goodman}}{{Foreman-Mackey} et~al.}{2013}]{2013PASP..125..306F}
{Foreman-Mackey} D.,  {Hogg} D.~W.,  {Lang} D.,   {Goodman} J.,  2013, \mn@doi
  [\pasp] {10.1086/670067}, \href
  {https://ui.adsabs.harvard.edu/abs/2013PASP..125..306F} {125, 306}

\bibitem[\protect\citeauthoryear{{Frank}, {King}  \& {Lasota}}{{Frank}
  et~al.}{1988}]{FrankKingLasota1988}
{Frank} J.,  {King} A.~R.,   {Lasota} J.~P.,  1988, \aap, \href
  {https://ui.adsabs.harvard.edu/abs/1988A&A...193..113F} {193, 113}

\bibitem[\protect\citeauthoryear{{G{\"a}nsicke}, {Fischer}, {Silvotti}  \& {de
  Martino}}{{G{\"a}nsicke} et~al.}{2001}]{2001A&A...372..557G}
{G{\"a}nsicke} B.~T.,  {Fischer} A.,  {Silvotti} R.,   {de Martino} D.,  2001,
  \mn@doi [\aap] {10.1051/0004-6361:20010522}, \href
  {https://ui.adsabs.harvard.edu/abs/2001A&A...372..557G} {372, 557}

\bibitem[\protect\citeauthoryear{{Giampapa}, {Rosner}, {Kashyap}, {Fleming},
  {Schmitt}  \& {Bookbinder}}{{Giampapa} et~al.}{1996}]{1996ApJ...463..707G}
{Giampapa} M.~S.,  {Rosner} R.,  {Kashyap} V.,  {Fleming} T.~A.,  {Schmitt}
  J.~H.~M.~M.,   {Bookbinder} J.~A.,  1996, \mn@doi [\apj] {10.1086/177284},
  \href {https://ui.adsabs.harvard.edu/abs/1996ApJ...463..707G} {463, 707}

\bibitem[\protect\citeauthoryear{{Ginzburg} \& {Quataert}}{{Ginzburg} \&
  {Quataert}}{2021}]{Ginzburg2021MNRAS}
{Ginzburg} S.,  {Quataert} E.,  2021, \mn@doi [\mnras]
  {10.1093/mnras/stab2170}, \href
  {https://ui.adsabs.harvard.edu/abs/2021MNRAS.507..475G} {507, 475}

\bibitem[\protect\citeauthoryear{{Hakala}, {Ramsay}, {Potter}, {Beardmore},
  {Buckley}  \& {Wynn}}{{Hakala} et~al.}{2019}]{2019MNRAS.486.2549H}
{Hakala} P.,  {Ramsay} G.,  {Potter} S.~B.,  {Beardmore} A.,  {Buckley} D.
  A.~H.,   {Wynn} G.,  2019, \mn@doi [\mnras] {10.1093/mnras/stz992}, \href
  {https://ui.adsabs.harvard.edu/abs/2019MNRAS.486.2549H} {486, 2549}

\bibitem[\protect\citeauthoryear{{Hameury} \& {King}}{{Hameury} \&
  {King}}{1988}]{1988MNRAS.235..433H}
{Hameury} J.~M.,  {King} A.~R.,  1988, \mn@doi [\mnras]
  {10.1093/mnras/235.2.433}, \href
  {https://ui.adsabs.harvard.edu/abs/1988MNRAS.235..433H} {235, 433}

\bibitem[\protect\citeauthoryear{{Hameury} \& {Lasota}}{{Hameury} \&
  {Lasota}}{2002}]{2002A&A...394..231H}
{Hameury} J.~M.,  {Lasota} J.~P.,  2002, \mn@doi [\aap]
  {10.1051/0004-6361:20021136}, \href
  {https://ui.adsabs.harvard.edu/abs/2002A&A...394..231H} {394, 231}

\bibitem[\protect\citeauthoryear{{Hartquist} \& {Williams}}{{Hartquist} \&
  {Williams}}{1989}]{Hartquist1989}
{Hartquist} T.~W.,  {Williams} D.~A.,  1989, \mn@doi [\mnras]
  {10.1093/mnras/241.3.417}, \href
  {https://ui.adsabs.harvard.edu/abs/1989MNRAS.241..417H} {241, 417}

\bibitem[\protect\citeauthoryear{{Hessman}, {G{\"a}nsicke}  \&
  {Mattei}}{{Hessman} et~al.}{2000}]{2000A&A...361..952H}
{Hessman} F.~V.,  {G{\"a}nsicke} B.~T.,   {Mattei} J.~A.,  2000, \aap, \href
  {https://ui.adsabs.harvard.edu/abs/2000A&A...361..952H} {361, 952}

\bibitem[\protect\citeauthoryear{{Hill} et~al.,}{{Hill}
  et~al.}{2022}]{2022AJ....163..246H}
{Hill} K.~L.,  et~al., 2022, \mn@doi [\aj] {10.3847/1538-3881/ac5a51}, \href
  {https://ui.adsabs.harvard.edu/abs/2022AJ....163..246H} {163, 246}

\bibitem[\protect\citeauthoryear{{Honeycutt} \& {Kafka}}{{Honeycutt} \&
  {Kafka}}{2004}]{2004AJ....128.1279H}
{Honeycutt} R.~K.,  {Kafka} S.,  2004, \mn@doi [\aj] {10.1086/422737}, \href
  {https://ui.adsabs.harvard.edu/abs/2004AJ....128.1279H} {128, 1279}

\bibitem[\protect\citeauthoryear{{Kafka} \& {Honeycutt}}{{Kafka} \&
  {Honeycutt}}{2003}]{2003AJ....125.2188K}
{Kafka} S.,  {Honeycutt} R.~K.,  2003, \mn@doi [\aj] {10.1086/368143}, \href
  {https://ui.adsabs.harvard.edu/abs/2003AJ....125.2188K} {125, 2188}

\bibitem[\protect\citeauthoryear{{Kafka} \& {Honeycutt}}{{Kafka} \&
  {Honeycutt}}{2005}]{2005AJ....130..742K}
{Kafka} S.,  {Honeycutt} R.~K.,  2005, \mn@doi [\aj] {10.1086/431793}, \href
  {https://ui.adsabs.harvard.edu/abs/2005AJ....130..742K} {130, 742}

\bibitem[\protect\citeauthoryear{{Kafka}, {Honeycutt}, {Howell}  \&
  {Harrison}}{{Kafka} et~al.}{2005}]{2005AJ....130.2852K}
{Kafka} S.,  {Honeycutt} R.~K.,  {Howell} S.~B.,   {Harrison} T.~E.,  2005,
  \mn@doi [\aj] {10.1086/497893}, \href
  {https://ui.adsabs.harvard.edu/abs/2005AJ....130.2852K} {130, 2852}

\bibitem[\protect\citeauthoryear{{Kirby}}{{Kirby}}{1995}]{1995PhST...59...59K}
{Kirby} K.~P.,  1995, \mn@doi [Physica Scripta Volume T]
  {10.1088/0031-8949/1995/T59/007}, \href
  {https://ui.adsabs.harvard.edu/abs/1995PhST...59...59K} {59, 59}

\bibitem[\protect\citeauthoryear{{Kovetz}, {Prialnik}  \& {Shara}}{{Kovetz}
  et~al.}{1988}]{Kovetz1988ApJ}
{Kovetz} A.,  {Prialnik} D.,   {Shara} M.~M.,  1988, \mn@doi [\apj]
  {10.1086/166053}, \href
  {https://ui.adsabs.harvard.edu/abs/1988ApJ...325..828K} {325, 828}

\bibitem[\protect\citeauthoryear{{Kuijpers} \& {Pringle}}{{Kuijpers} \&
  {Pringle}}{1982}]{1982A&A...114L...4K}
{Kuijpers} J.,  {Pringle} J.~E.,  1982, \aap, \href
  {https://ui.adsabs.harvard.edu/abs/1982A&A...114L...4K} {114, L4}

\bibitem[\protect\citeauthoryear{{Leach}, {Hessman}, {King}, {Stehle}  \&
  {Mattei}}{{Leach} et~al.}{1999}]{1999MNRAS.305..225L}
{Leach} R.,  {Hessman} F.~V.,  {King} A.~R.,  {Stehle} R.,   {Mattei} J.,
  1999, \mn@doi [\mnras] {10.1046/j.1365-8711.1999.02450.x}, \href
  {https://ui.adsabs.harvard.edu/abs/1999MNRAS.305..225L} {305, 225}

\bibitem[\protect\citeauthoryear{{Lightkurve Collaboration}
  et~al.,}{{Lightkurve Collaboration} et~al.}{2018}]{2018ascl.soft12013L}
{Lightkurve Collaboration} et~al., 2018, {Lightkurve: Kepler and TESS time
  series analysis in Python} (\mn@eprint {ascl} {1812.013})

\bibitem[\protect\citeauthoryear{{Livio} \& {Pringle}}{{Livio} \&
  {Pringle}}{1994}]{1994ApJ...427..956L}
{Livio} M.,  {Pringle} J.~E.,  1994, \mn@doi [\apj] {10.1086/174202}, \href
  {https://ui.adsabs.harvard.edu/abs/1994ApJ...427..956L} {427, 956}

\bibitem[\protect\citeauthoryear{{Lubow} \& {Shu}}{{Lubow} \&
  {Shu}}{1975}]{Lubow1975ApJ}
{Lubow} S.~H.,  {Shu} F.~H.,  1975, \mn@doi [\apj] {10.1086/153614}, \href
  {https://ui.adsabs.harvard.edu/abs/1975ApJ...198..383L} {198, 383}

\bibitem[\protect\citeauthoryear{{Mason} \& {Santana}}{{Mason} \&
  {Santana}}{2015}]{2015gacv.workE..16M}
{Mason} P.~A.,  {Santana} J.~B.,  2015, in The Golden Age of Cataclysmic
  Variables and Related Objects - III (Golden2015). p.~16

\bibitem[\protect\citeauthoryear{{Mason}, {Middleditch}, {Cordova}, {Jensen},
  {Reichert}, {Murdin}, {Clark}  \& {Bowyer}}{{Mason}
  et~al.}{1983}]{1983ApJ...264..575M}
{Mason} K.~O.,  {Middleditch} J.,  {Cordova} F.~A.,  {Jensen} K.~A.,
  {Reichert} G.,  {Murdin} P.~G.,  {Clark} D.,   {Bowyer} S.,  1983, \mn@doi
  [\apj] {10.1086/160627}, \href
  {https://ui.adsabs.harvard.edu/abs/1983ApJ...264..575M} {264, 575}

\bibitem[\protect\citeauthoryear{{Meyer} \& {Meyer-Hofmeister}}{{Meyer} \&
  {Meyer-Hofmeister}}{1983}]{Meyer1983A&A}
{Meyer} F.,  {Meyer-Hofmeister} E.,  1983, \aap, \href
  {https://ui.adsabs.harvard.edu/abs/1983A&A...121...29M} {121, 29}

\bibitem[\protect\citeauthoryear{{Middleditch}, {Imamura}, {Wolff}  \&
  {Steiman-Cameron}}{{Middleditch} et~al.}{1991}]{1991ApJ...382..315M}
{Middleditch} J.,  {Imamura} J.~N.,  {Wolff} M.~T.,   {Steiman-Cameron} T.~Y.,
  1991, \mn@doi [\apj] {10.1086/170718}, \href
  {https://ui.adsabs.harvard.edu/abs/1991ApJ...382..315M} {382, 315}

\bibitem[\protect\citeauthoryear{{Mouchet} et~al.,}{{Mouchet}
  et~al.}{2017}]{2017A&A...600A..53M}
{Mouchet} M.,  et~al., 2017, \mn@doi [\aap] {10.1051/0004-6361/201630166},
  \href {https://ui.adsabs.harvard.edu/abs/2017A&A...600A..53M} {600, A53}

\bibitem[\protect\citeauthoryear{Newville et~al.,}{Newville
  et~al.}{2021}]{matt_newville_2021_5570790}
Newville M.,  et~al., 2021, lmfit/lmfit-py: 1.0.3,
  \mn@doi{10.5281/zenodo.5570790}, \url
  {https://doi.org/10.5281/zenodo.5570790}

\bibitem[\protect\citeauthoryear{{Nousek} et~al.,}{{Nousek}
  et~al.}{1984}]{1984ApJ...277..682N}
{Nousek} J.~A.,  et~al., 1984, \mn@doi [\apj] {10.1086/161739}, \href
  {https://ui.adsabs.harvard.edu/abs/1984ApJ...277..682N} {277, 682}

\bibitem[\protect\citeauthoryear{{Olivero}}{{Olivero}}{1977}]{1977JQSRT..17..233O}
{Olivero} J.,  1977, \mn@doi [\jqsrt] {10.1016/0022-4073(77)90161-3}, \href
  {https://ui.adsabs.harvard.edu/abs/1977JQSRT..17..233O} {17, 233}

\bibitem[\protect\citeauthoryear{{Osaki}}{{Osaki}}{1985}]{Osaki1985A&A}
{Osaki} Y.,  1985, \aap, \href
  {https://ui.adsabs.harvard.edu/abs/1985A&A...144..369O} {144, 369}

\bibitem[\protect\citeauthoryear{{Plavec}, {Ulrich}  \& {Polidan}}{{Plavec}
  et~al.}{1973}]{Plavec1973PASP}
{Plavec} M.,  {Ulrich} R.~K.,   {Polidan} R.~S.,  1973, \mn@doi [\pasp]
  {10.1086/129546}, \href
  {https://ui.adsabs.harvard.edu/abs/1973PASP...85..769P} {85, 769}

\bibitem[\protect\citeauthoryear{{Potter}, {Augusteijn}  \& {Tappert}}{{Potter}
  et~al.}{2005}]{Potter2005MNRAS}
{Potter} S.~B.,  {Augusteijn} T.,   {Tappert} C.,  2005, \mn@doi [\mnras]
  {10.1111/j.1365-2966.2005.09569.x}, \href
  {https://ui.adsabs.harvard.edu/abs/2005MNRAS.364..565P} {364, 565}

\bibitem[\protect\citeauthoryear{Ramsay}{Ramsay}{2021}]{Ramsay}
Ramsay G.,  2021, Private communication

\bibitem[\protect\citeauthoryear{{Ramsay}, {Cropper}  \& {Mason}}{{Ramsay}
  et~al.}{1995}]{1995MNRAS.276.1382R}
{Ramsay} G.,  {Cropper} M.,   {Mason} K.~O.,  1995, \mn@doi [\mnras]
  {10.1093/mnras/276.4.1382}, \href
  {https://ui.adsabs.harvard.edu/abs/1995MNRAS.276.1382R} {276, 1382}

\bibitem[\protect\citeauthoryear{{Ramsay}, {Hakala}  \& {Wood}}{{Ramsay}
  et~al.}{2021}]{Ramsay2021}
{Ramsay} G.,  {Hakala} P.,   {Wood} M.~A.,  2021, \mn@doi [\mnras]
  {10.1093/mnras/stab1140}, \href
  {https://ui.adsabs.harvard.edu/abs/2021MNRAS.504.4072R} {504, 4072}

\bibitem[\protect\citeauthoryear{{Ricker} et~al.,}{{Ricker}
  et~al.}{2015}]{2015JATIS...1a4003R}
{Ricker} G.~R.,  et~al., 2015, \mn@doi [Journal of Astronomical Telescopes,
  Instruments, and Systems] {10.1117/1.JATIS.1.1.014003}, \href
  {https://ui.adsabs.harvard.edu/abs/2015JATIS...1a4003R} {1, 014003}

\bibitem[\protect\citeauthoryear{{Ritter}}{{Ritter}}{1988}]{Ritter1988A&A}
{Ritter} H.,  1988, \aap, \href
  {https://ui.adsabs.harvard.edu/abs/1988A&A...202...93R} {202, 93}

\bibitem[\protect\citeauthoryear{{Romero-Colmenero}, {Potter}, {Buckley},
  {Barrett}  \& {Vrielmann}}{{Romero-Colmenero}
  et~al.}{2003}]{2003MNRAS.339..685R}
{Romero-Colmenero} E.,  {Potter} S.~B.,  {Buckley} D.~A.~H.,  {Barrett} P.~E.,
   {Vrielmann} S.,  2003, \mn@doi [\mnras] {10.1046/j.1365-8711.2003.06209.x},
  \href {https://ui.adsabs.harvard.edu/abs/2003MNRAS.339..685R} {339, 685}

\bibitem[\protect\citeauthoryear{{Sarna}}{{Sarna}}{1990}]{Sarna1990A&A}
{Sarna} M.~J.,  1990, \aap, \href
  {https://ui.adsabs.harvard.edu/abs/1990A&A...239..163S} {239, 163}

\bibitem[\protect\citeauthoryear{{Schwarz}, {Greiner}, {Tovmassian}, {Zharikov}
   \& {Wenzel}}{{Schwarz} et~al.}{2002}]{2002A&A...392..505S}
{Schwarz} R.,  {Greiner} J.,  {Tovmassian} G.~H.,  {Zharikov} S.~V.,   {Wenzel}
  W.,  2002, \mn@doi [\aap] {10.1051/0004-6361:20021193}, \href
  {https://ui.adsabs.harvard.edu/abs/2002A&A...392..505S} {392, 505}

\bibitem[\protect\citeauthoryear{{Schwope}, {Beuermann}  \& {Thomas}}{{Schwope}
  et~al.}{1990}]{1990A&A...230..120S}
{Schwope} A.~D.,  {Beuermann} K.,   {Thomas} H.~C.,  1990, \aap, \href
  {https://ui.adsabs.harvard.edu/abs/1990A&A...230..120S} {230, 120}

\bibitem[\protect\citeauthoryear{{Schwope}, {Thomas}, {Beuermann}  \&
  {Naundorf}}{{Schwope} et~al.}{1991}]{1991A&A...244..373S}
{Schwope} A.~D.,  {Thomas} H.~C.,  {Beuermann} K.,   {Naundorf} C.~E.,  1991,
  \aap, \href {https://ui.adsabs.harvard.edu/abs/1991A&A...244..373S} {244,
  373}

\bibitem[\protect\citeauthoryear{{Schwope}, {Thomas}, {Beuermann}  \&
  {Reinsch}}{{Schwope} et~al.}{1993}]{1993A&A...267..103S}
{Schwope} A.~D.,  {Thomas} H.~C.,  {Beuermann} K.,   {Reinsch} K.,  1993, \aap,
  \href {https://ui.adsabs.harvard.edu/abs/1993A&A...267..103S} {267, 103}

\bibitem[\protect\citeauthoryear{{Schwope}, {Catal{\'a}n}, {Beuermann},
  {Metzner}, {Smith}  \& {Steeghs}}{{Schwope}
  et~al.}{2000}]{2000MNRAS.313..533S}
{Schwope} A.~D.,  {Catal{\'a}n} M.~S.,  {Beuermann} K.,  {Metzner} A.,  {Smith}
  R.~C.,   {Steeghs} D.,  2000, \mn@doi [\mnras]
  {10.1046/j.1365-8711.2000.03240.x}, \href
  {https://ui.adsabs.harvard.edu/abs/2000MNRAS.313..533S} {313, 533}

\bibitem[\protect\citeauthoryear{{Schwope}, {Brunner}, {Buckley}, {Greiner},
  {Heyden}, {Neizvestny}, {Potter}  \& {Schwarz}}{{Schwope}
  et~al.}{2002}]{2002A&A...396..895S}
{Schwope} A.~D.,  {Brunner} H.,  {Buckley} D.,  {Greiner} J.,  {Heyden} K.
  v.~d.,  {Neizvestny} S.,  {Potter} S.,   {Schwarz} R.,  2002, \mn@doi [\aap]
  {10.1051/0004-6361:20021386}, \href
  {https://ui.adsabs.harvard.edu/abs/2002A&A...396..895S} {396, 895}

\bibitem[\protect\citeauthoryear{{Schwope}, {Worpel}, {Traulsen}  \&
  {Sablowski}}{{Schwope} et~al.}{2020}]{2020A&A...642A.134S}
{Schwope} A.~D.,  {Worpel} H.,  {Traulsen} I.,   {Sablowski} D.,  2020, \mn@doi
  [\aap] {10.1051/0004-6361/202037714}, \href
  {https://ui.adsabs.harvard.edu/abs/2020A&A...642A.134S} {642, A134}

\bibitem[\protect\citeauthoryear{{\v Simon}}{{\v
  Simon}}{2016}]{2016MNRAS.463.1342S}
{\v Simon} V.,  2016, \mn@doi [\mnras] {10.1093/mnras/stw1964}, \href
  {https://ui.adsabs.harvard.edu/abs/2016MNRAS.463.1342S} {463, 1342}

\bibitem[\protect\citeauthoryear{{Spitzer}}{{Spitzer}}{1978}]{Spitzer1978}
{Spitzer} L.,  1978, {Physical processes in the interstellar medium},
  \mn@doi{10.1002/9783527617722.
}

\bibitem[\protect\citeauthoryear{{Szkody} et~al.,}{{Szkody}
  et~al.}{2002}]{2002AJ....123..430S}
{Szkody} P.,  et~al., 2002, \mn@doi [\aj] {10.1086/324734}, \href
  {https://ui.adsabs.harvard.edu/abs/2002AJ....123..430S} {123, 430}

\bibitem[\protect\citeauthoryear{{Tapia}}{{Tapia}}{1977}]{Tapia77}
{Tapia} S.,  1977, \mn@doi [\apjl] {10.1086/182390}, \href
  {https://ui.adsabs.harvard.edu/abs/1977ApJ...212L.125T} {212, L125}

\bibitem[\protect\citeauthoryear{{Wang} \& {Zhong}}{{Wang} \&
  {Zhong}}{2018}]{Wang2018A&A}
{Wang} J.,  {Zhong} Z.,  2018, \mn@doi [\aap] {10.1051/0004-6361/201834109},
  \href {https://ui.adsabs.harvard.edu/abs/2018A&A...619L...1W} {619, L1}

\bibitem[\protect\citeauthoryear{Warner}{Warner}{1995}]{warner_1995}
Warner B.,  1995, Cataclysmic Variable Stars.
Cambridge Astrophysics, Cambridge University Press,
  \mn@doi{10.1017/CBO9780511586491}

\bibitem[\protect\citeauthoryear{{Wickramasinghe} \& {Wu}}{{Wickramasinghe} \&
  {Wu}}{1991}]{1991MNRAS.253P..11W}
{Wickramasinghe} D.~T.,  {Wu} K.,  1991, \mn@doi [\mnras]
  {10.1093/mnras/253.1.11P}, \href
  {https://ui.adsabs.harvard.edu/abs/1991MNRAS.253P..11W} {253, 11P}

\bibitem[\protect\citeauthoryear{{Wickramasinghe}, {Ferrario}  \&
  {Bailey}}{{Wickramasinghe} et~al.}{1989}]{1989ApJ...342L..35W}
{Wickramasinghe} D.~T.,  {Ferrario} L.,   {Bailey} J.,  1989, \mn@doi [\apjl]
  {10.1086/185478}, \href
  {https://ui.adsabs.harvard.edu/abs/1989ApJ...342L..35W} {342, L35}

\bibitem[\protect\citeauthoryear{{Wickramasinghe}, {Bailey}, {Meggitt},
  {Ferrario}, {Hough}  \& {Tuohy}}{{Wickramasinghe}
  et~al.}{1991}]{1991MNRAS.251...28W}
{Wickramasinghe} D.~T.,  {Bailey} J.,  {Meggitt} S.~M.~A.,  {Ferrario} L.,
  {Hough} J.,   {Tuohy} I.~R.,  1991, \mn@doi [\mnras]
  {10.1093/mnras/251.1.28}, \href
  {https://ui.adsabs.harvard.edu/abs/1991MNRAS.251...28W} {251, 28}

\bibitem[\protect\citeauthoryear{{Wu}}{{Wu}}{2000}]{2000SSRv...93..611W}
{Wu} K.,  2000, \mn@doi [\ssr] {10.1023/A:1026522914125}, \href
  {https://ui.adsabs.harvard.edu/abs/2000SSRv...93..611W} {93, 611}

\bibitem[\protect\citeauthoryear{{Wu} \& {Kiss}}{{Wu} \&
  {Kiss}}{2008}]{2008A&A...481..433W}
{Wu} K.,  {Kiss} L.~L.,  2008, \mn@doi [\aap] {10.1051/0004-6361:20078556},
  \href {https://ui.adsabs.harvard.edu/abs/2008A&A...481..433W} {481, 433}

\bibitem[\protect\citeauthoryear{{Wu} \& {Wickramasinghe}}{{Wu} \&
  {Wickramasinghe}}{1993}]{1993MNRAS.260..141W}
{Wu} K.,  {Wickramasinghe} D.~T.,  1993, \mn@doi [\mnras]
  {10.1093/mnras/260.1.141}, \href
  {https://ui.adsabs.harvard.edu/abs/1993MNRAS.260..141W} {260, 141}

\bibitem[\protect\citeauthoryear{{Wu}, {Wickramasinghe}  \& {Warner}}{{Wu}
  et~al.}{1995}]{Wu1995PASA}
{Wu} K.,  {Wickramasinghe} D.~T.,   {Warner} B.,  1995, \mn@doi [\pasa]
  {10.1017/S132335800002004X}, \href
  {https://ui.adsabs.harvard.edu/abs/1995PASA...12...60W} {12, 60}

\bibitem[\protect\citeauthoryear{{Zubareva}, {Pavlenko}, {Andreev}, {Antipin},
  {Samus'}  \& {Sergeev}}{{Zubareva} et~al.}{2011}]{2011ARep...55..224Z}
{Zubareva} A.~M.,  {Pavlenko} E.~P.,  {Andreev} M.~V.,  {Antipin} S.~V.,
  {Samus'} N.~N.,   {Sergeev} A.~V.,  2011, \mn@doi [Astronomy Reports]
  {10.1134/S1063772911030097}, \href
  {https://ui.adsabs.harvard.edu/abs/2011ARep...55..224Z} {55, 224}

\makeatother
\end{thebibliography}

\appendix

\section{Magnetic confinement of outflow of cool material 
  at the L1 point}\label{appx:ambi}

{{The donor stars in Polars are mostly low-mass M stars. Although they are considered cool in terms of the spectral temperature of their photospheric emission, they are magnetically active and have a corona. The gas in their corona is ionised and it is also collisional in nature \citep{1996ApJ...463..707G}. The interaction of the magnetised plasma in the corona results in the ionised gas mixing with the cooler gas at the stellar surface that gives rise to the optical/IR photospheric emission of the star. While the charged particles are strongly coupled with the magnetic field, neutral gas with temperature as low as 2000$-$3000~K, indicated by the photospheric emission from M star would, in principle, not be confined magnetically. However, because of collisions between the neutral atoms and the charged particles, their mobility would be restricted by the magnetic field. The neutral gas therefore does not free stream across the magnetic field but undergoes an ambipolar diffusion, which may be described by the following relation:  
\begin{align}  
 \alpha_{\rm in} n_{\rm i} \rho v_{\rm d} & \approx \frac{B^2}{4\pi L}  
\end{align} 
\citep[see][]{Spitzer1978}, where $n_{\rm i}$ is the number density of ions, $\rho$ is the mass density of the medium, $v_{\rm d}$ is the diffusion speed, $L$ is the characteristic linear size of the medium where the neutral gas diffuses across. The collisional rate between the neutral and charged particles, $\alpha_{\rm in}$, in astrophysical systems generally have values about $10^{-9}{\rm cm}^3 {\rm s}^{-1}$ \citep[see][]{1995PhST...59...59K,Hartquist1989}.}}

{{
We may express the number density of ions in terms of the total number density of atomic particles as $n_{\rm i} = \chi n_{\rm p}$, with $\chi$ as a parameter effectively indicating the ionisation faction. Magnetic confinement requires that 
\begin{align} 
  \frac{B^2}{8 \pi \rho {c_{\rm s}}^2}&  > 1 \ , 
\end{align} 
where $c_{\rm s}$ is the local isothermal sound speed of the particles confined by the magnetic field. It follows that the criterion for magnetic confinement of ions and particles in the presence of ambipolar diffusion is given by 
\begin{align} 
   \frac{\alpha_{\rm in}\chi  n_{\rm p}  v_{\rm d} L}{2 {c_{\rm s,n}}^2 \left[(1-\chi) + \chi (c_{\rm s,i}/c_{\rm s,n})^2   \right] }    & > 1  \ , 
\end{align}  
where $c_{\rm s,n}$ and $c_{\rm s,i}$ are the thermal sound speeds of the neutral particles and the ions respectively.  
}}
 
{{ 
For mass transfer in low-mass semi-detached binaries, the outflow from the donor stars at the L1 point is transonic \citep[see][]{Lubow1975ApJ}. With this condition, we have ${\rm Max}(v_{\rm d}) \sim |c_{\rm s,n} - c_{\rm s,i} | \approx  c_{\rm s,i}$, and the confinement condition becomes 
 \begin{align}
    \chi  n_{\rm p} &  >
    \frac{2\; \! {c_{\rm s,n}}^2}{\alpha_{\rm in} L c_{\rm s, i}}  \ . 
\end{align}
 }}
 
{{Assuming $L \sim 10^9 {\rm cm}$, $c_{\rm s, n} \approx 10^5{\rm cm}\;\!{\rm s}^{-1}$ (corresponding to a thermal temperature of $2000~{\rm K}$) and $c_{\rm s, i} \approx 2 \times 10^7{\rm cm}\;\!{\rm s}^{-1}$ (corresponding to a thermal temperature of $10^{6}~{\rm K}$) gives  
 \begin{align} 
 \chi  n_{\rm p}  > 10^3{\rm cm}^{-3}  \ .    
 \end{align}  
Thus, the presence of a small amount of ions would be able to hold the neutral gas at the L1 point. To violate the confinement condition, it requires a very low matter density and a low ionisation such that $\chi  n_{\rm p} < 10^{3} {\rm cm}^{-3}$, or the presence of a strong external force that has, however, built up to counterbalance the magnetic stress force. Alternatively, the condition does not hold if the transonic condition for mass transfer in low-mass semi-detached systems as that derived by \cite{Lubow1975ApJ} is invalid.  
} }

\section{Complete List of ZTF Sources}

\begin{table*}
\caption{ZTF sources considered. A number of sources show no values, this is because they were excluded from analysis as they had fewer than 25 data points in both filters. The state changes column relates to the presence of short duration, long duration state changes, both or none.}\label{appx1}
\begin{tabular}{p{3.5cm}>{\centering\arraybackslash}p{1cm}>{\centering\arraybackslash}p{1cm}>{\centering\arraybackslash}p{1cm}>{\centering\arraybackslash}p{1cm}>{\centering\arraybackslash}p{1cm}>{\centering\arraybackslash}p{1cm}>{\centering\arraybackslash}p{1.8cm}>{\centering\arraybackslash}p{1.8cm}}
\hline
Source & g mean & g min & g max & r mean & r min & r max & No. Data Points & State Changes \\
\hline
2XMM J183251.4-100106 & 17.8 & 17.94 & 17.58 & 17.47 & 20.5 & 15.57 & 1569 & None \\ \
2XMMp J1313223.4+173659 & 19.09 & 20.48 & 18.01 & 19.03 & 20.41 & 17.92 & 2179 & None \\ \
AM Her & 14.26 & 15.47 & 12.79 & 13.99 & 15.33 & 12.28 & 1843 & Both \\ \
AN UMa & 17.21 & 19.17 & 15.88 & 17.09 & 18.86 & 15.85 & 843 & Short \\ \
AP CrB & 17.4 & 17.73 & 16.14 & 17.42 & 17.78 & 16.2 & 543 & Short \\ \
AR UMa & 16.36 & 16.41 & 16.31 & 16.59 & 16.7 & 16.51 & 118 & None \\ \
BS Tri & 18.2 & 20.5 & 17.03 & 17.84 & 20.24 & 16.64 & 924 & Short \\ \
BY Cam & 15.38 & 16.2 & 14.4 & 14.97 & 16.07 & 13.71 & 726 & None \\ \
CSS0357+10 & 19.34 & 20.39 & 18.09 & 18.49 & 19.46 & 17.3 & 91 & None \\ \
CSS2335+12 & 19.17 & 20.47 & 17.66 & 18.83 & 20.18 & 17.59 & 579 & None \\ \
        DDE23 & 19.19 & 19.97 & 17.99 & 18.77 & 19.89 & 17.42 & 3808 & None \\ \
        DP Leo & 18.76 & 19.99 & 18.19 & 18.62 & 20.2 & 18.03 & 267 & None \\ \
        EF Eri & 18.13 & 18.3 & 17.97 & 18.28 & 18.41 & 18.16 & 145 & None \\ \
        EG Lyn & 18.62 & 19.09 & 18.01 & 18.39 & 18.87 & 17.76 & 83 & None \\ \
        EK UMa & 18.96 & 20.43 & 17.74 & 18.94 & 20.18 & 17.88 & 948 & None \\ \
        EP Dra & 18.35 & 20.43 & 17.29 & 18.23 & 20.43 & 16.84 & 1548 & None \\ \
        EQ Cet & 18.02 & 19.0 & 16.79 & 17.94 & 18.96 & 16.85 & 227 & None \\ \
        EU Cnc & ——— & ——— & ——— & ——— & ——— & ——— & 10 & ——— \\ \
        EU UMa & 18.4 & 19.32 & 17.04 & 18.39 & 20.5 & 16.72 & 1512 & Short \\ \
        EV UMa & 17.67 & 20.5 & 15.89 & 17.85 & 20.47 & 16.23 & 1246 & Long \\ \
        FH UMa & ——— & ——— & ——— & ——— & ——— & ——— & 38 & ——— \\ \
        FL Cet & 19.15 & 20.12 & 17.01 & 19.01 & 19.96 & 16.37 & 437 & Short \\ \
        GG Leo & 18.19 & 19.34 & 16.39 & 17.87 & 19.25 & 16.12 & 374 & None \\ \
        HS Cam & 19.98 & 20.5 & 18.14 & 19.86 & 20.5 & 17.83 & 798 & Short \\ \
        HU Aqr & 16.24 & 18.65 & 14.63 & 17.09 & 20.49 & 14.58 & 412 & None \\ \
        HY Eri & 19.59 & 20.5 & 18.79 & 19.41 & 20.49 & 18.4 & 1072 & None \\ \
        IW Eri & 18.68 & 19.87 & 16.28 & 18.65 & 19.92 & 16.27 & 264 & Long \\ \
        LW Cam & 18.01 & 20.49 & 15.43 & 18.63 & 20.5 & 15.22 & 760 & Long \\ \
        MN Hya & 18.02 & 19.79 & 16.58 & 18.08 & 19.72 & 16.72 & 51 & None \\ \
        MQ Dra & 18.69 & 19.14 & 18.24 & 17.49 & 18.13 & 16.93 & 1689 & None \\ \
        MR Ser & ——— & ——— & ——— & ——— & ——— & ——— & 27 & ——— \\ \
        MT Dra & 18.12 & 20.49 & 17.01 & 17.6 & 20.48 & 16.44 & 1959 & Short \\ \
        PBC J0706.7+0327 & 17.14 & 17.66 & 16.61 & 17.11 & 17.73 & 16.39 & 1154 & None \\ \
        QQ Vul & 15.05 & 17.05 & 14.37 & 14.87 & 16.37 & 14.16 & 1109 & None \\ \
        RX J0325-08 & 17.59 & 18.54 & 15.48 & 17.2 & 18.48 & 15.01 & 418 & None \\ \
        RX J0502.8+1624 & 19.05 & 19.32 & 18.44 & 19.0 & 19.29 & 17.78 & 545 & None \\ \
        RX J0524+4244 & 17.35 & 18.05 & 16.41 & 17.36 & 18.13 & 16.39 & 818 & None \\ \
        RX J0600-2709 & 19.72 & 20.2 & 19.2 & 19.59 & 20.41 & 18.9 & 598 & None \\ \
        RX J0649-0737 & 17.24 & 18.83 & 16.24 & 16.75 & 17.9 & 16.25 & 398 & None \\ \
        RX J0749-0549 & 18.73 & 20.37 & 17.31 & 19.4 & 20.3 & 16.98 & 659 & Long \\ \
        RX J0953+1458 & 18.86 & 20.45 & 16.56 & 18.33 & 20.43 & 16.42 & 408 & None \\ \
        RX J1002-19 & 18.43 & 19.63 & 17.07 & 18.81 & 20.32 & 17.22 & 176 & None \\ \
        RX J1007-20 & 18.5 & 20.29 & 16.77 & 18.15 & 19.54 & 16.53 & 183 & None \\ \
        RX J1610+03 & 17.49 & 19.07 & 16.24 & 16.98 & 18.65 & 15.33 & 403 & Short \\ \
        RX J2015.6+3711 & 18.69 & 19.07 & 16.54 & 17.51 & 17.87 & 15.85 & 1258 & None \\ \
        RX J2316-05 & 19.55 & 20.2 & 17.85 & 18.98 & 19.8 & 16.91 & 326 & None \\ \
        SDSS J032855+052254 & 19.36 & 20.45 & 17.5 & 19.13 & 20.49 & 17.48 & 317 & None \\ \
        SDSS J072910+365838 & 20.14 & 20.49 & 19.19 & 19.91 & 20.49 & 18.87 & 779 & None \\ \
        SDSS J075240+362823 & 16.91 & 16.97 & 16.78 & 16.34 & 16.4 & 16.29 & 816 & None \\ \
        SDSS J085414+390537 & 18.59 & 19.93 & 15.61 & 18.61 & 20.31 & 15.88 & 772 & None \\ \
        SDSS J085909+053654 & 18.48 & 19.21 & 17.67 & 18.15 & 19.0 & 16.5 & 654 & None \\ \
        SDSS J092122+2038571 & ——— & ——— & ——— & ——— & ——— & ——— & 124 & ——— \\ \
        SDSS J103100+202832 & 18.12 & 20.48 & 16.44 & 18.26 & 20.14 & 16.79 & 380 & Long \\ \
        SDSS J121209+013627 & 17.96 & 18.07 & 17.82 & 18.09 & 18.21 & 17.97 & 497 & None \\ \
        SDSS J1333092+1437069 & 20.01 & 20.49 & 19.39 & 19.78 & 20.46 & 18.52 & 71 & None \\ \
        SDSS J142256-022108 & 19.4 & 20.45 & 18.56 & 19.23 & 20.32 & 18.38 & 276 & None \\ \
        SDSS J154104+360252 & 18.03 & 20.2 & 15.99 & 17.76 & 20.2 & 15.52 & 1466 & Long \\ \
        SDSS J170053+4000357 & 18.84 & 20.46 & 16.49 & 18.47 & 20.48 & 16.34 & 1743 & None \\ \
        SDSS J204827+005008 & 19.48 & 19.95 & 18.53 & 18.74 & 19.13 & 18.22 & 355 & None \\ \
        ST LMi & 16.61 & 17.66 & 15.15 & 16.31 & 17.63 & 14.38 & 507 & Short \\ \
        Swift J2341.0+7645 & 18.35 & 20.0 & 16.96 & 17.59 & 19.04 & 16.3 & 658 & Long \\ \

        \end{tabular}
\end{table*}

\begin{table*}
\contcaption{}
\begin{tabular}{p{3.5cm}>{\centering\arraybackslash}p{1cm}>{\centering\arraybackslash}p{1cm}>{\centering\arraybackslash}p{1cm}>{\centering\arraybackslash}p{1cm}>{\centering\arraybackslash}p{1cm}>{\centering\arraybackslash}p{1cm}>{\centering\arraybackslash}p{1.8cm}>{\centering\arraybackslash}p{1.8cm}}
        UZ For & 16.95 & 20.43 & 15.77 & 17.0 & 20.3 & 15.84 & 217 & None \\ 
        V1007 Her & ——— & ——— & ——— & ——— & ——— & ——— & 2 & ——— \\ 
        V1309 Ori & 16.21 & 18.07 & 15.08 & 15.83 & 20.49 & 15.04 & 883 & None \\ 
        V1432 Aql & 15.91 & 20.05 & 14.78 & 15.89 & 19.0 & 14.53 & 619 & None \\ 
        V1500 Cyg & 19.05 & 19.75 & 18.4 & 18.29 & 19.79 & 17.76 & 1778 & None \\ 
        V2301 Oph & 17.22 & 20.32 & 16.12 & 17.38 & 20.03 & 15.91 & 1723 & Short \\ 
        V388 Peg & 19.19 & 20.5 & 17.32 & 19.05 & 20.48 & 16.98 & 434 & Both \\ 
        V884 Her & 15.98 & 16.8 & 14.41 & 15.8 & 16.99 & 14.28 & 1497 & Short \\ 
        VY For & 18.01 & 19.66 & 16.6 & 17.78 & 19.45 & 16.58 & 184 & Long \\ 
        WX LMi & 17.57 & 17.83 & 17.27 & 16.89 & 17.08 & 16.57 & 279 & None \\ 
        \hline
\end{tabular}
\end{table*}

%\begin{table*}
%\contcaption{}
%\begin{tabular}{lcccccccc}
%2XMMpJ1313223.4+173659 & 19.09 & 20.48 & 18.01 & 19.03 & 20.41 & 17.92 & 2179 & None \\
%RXJ0325-08 & 17.59 & 18.54 & 15.48 & 17.2 & 18.48 & 15.01 & 418 & None \\
%V2301 Oph & 17.22 & 20.32 & 16.12 & 17.38 & 20.03 & 15.91 & 1723 & Short \\
%FH UMa & ——— & ——— & ——— & ——— & ——— & ——— & 38 & ——— \\
%EQ Cet & 18.02 & 19.0 & 16.79 & 17.94 & 18.96 & 16.85 & 227 & None \\
%RXJ0649-0737 & 17.24 & 18.83 & 16.24 & 16.75 & 17.9 & 16.25 & 398 & None \\
%BS Tri & 18.2 & 20.5 & 17.03 & 17.84 & 20.24 & 16.64 & 924 & Short \\
%QQ Vul & 15.05 & 17.05 & 14.37 & 14.87 & 16.37 & 14.16 & 1109 & None \\
%V884 Her & 15.98 & 16.8 & 14.41 & 15.8 & 16.99 & 14.28 & 1497 & Short \\
%LW Cam & 18.01 & 20.49 & 15.43 & 18.63 & 20.5 & 15.22 & 760 & Long \\
%VY For & 18.01 & 19.66 & 16.6 & 17.78 & 19.45 & 16.58 & 184 & Long \\
%SDSSJ072910+365838 & 20.14 & 20.49 & 19.19 & 19.91 & 20.49 & 18.87 & 779 & None \\
%\hline
%\end{tabular}
%\end{table*}

% Don't change these lines
\bsp	% typesetting comment
\label{lastpage}
\end{document}